\documentclass[11pt]{article}
\usepackage{amssymb,latexsym, amsmath}
\newcommand{\bfi}{\bfseries\itshape}

\makeatletter
\@addtoreset{figure}{section}
\def\thefigure{\thesection.\@arabic\c@figure}
\def\fps@figure{h, t}
\@addtoreset{table}{bsection}
\def\thetable{\thesection.\@arabic\c@table}
\def\fps@table{h, t}
\@addtoreset{equation}{section}

\makeatother

\newtheorem{thm}{Theorem}[section]

\newtheorem{lem}[thm]{Lemma}
\newtheorem{cor}[thm]{Corollary}

\begin{document}

\title{Lagrangian Reduction, the Euler--Poincar\'{e}
Equations, and Semidirect Products}
\author{Hern\'an Cendra\\
Universidad Nacional del Sur\\
8000 Bahia Blanca, Argentina\\
{\footnotesize uscendra@criba.edu.ar}
\and
Darryl D. Holm
\\Theoretical Division and Center for Nonlinear Studies
\\Los Alamos National
Laboratory, MS B284
\\ Los Alamos, NM 87545
\\ {\footnotesize dholm@lanl.gov}
\and
Jerrold E. Marsden\thanks{Research partially supported by
NSF grant DMS 96--33161 and DOE contract
DE--FG0395--ER25251}\\Control and Dynamical Systems\\
California Institute of Technology 107-81\\ Pasadena, CA 91125
\\ {\footnotesize  marsden@cds.caltech.edu}
\and
Tudor S. Ratiu
\thanks{Research partially supported by NSF Grant DMS-9503273 and DOE
contract DE-FG03-95ER25245-A000.}
\\Department of Mathematics
\\University of California, Santa Cruz, CA 95064
\\ {\footnotesize  ratiu@math.ucsc.edu} }
\date{February 1997; this version, October 9, 1997 \\
\small{\it To appear in the AMS Arnold Volume II}}

\maketitle

\begin{abstract}
There is a well developed and useful theory of Hamiltonian reduction
for semidirect products, which applies to examples such as the heavy
top, compressible fluids and MHD, which are governed by Lie-Poisson
type equations. In this paper we study the Lagrangian analogue of this
process and link it with the general theory of Lagrangian reduction;
that is the reduction of variational principles. These reduced
variational principles are interesting in their own right since they
involve constraints on the allowed variations, analogous to what one
finds in the theory of nonholonomic systems with the La\-gran\-ge
d'Al\-em\-bert principle. In addition, the abstract theorems about
circulation, what we call the Kelvin-Noether theorem, are given.
\end{abstract}

\tableofcontents

\section{Introduction}

The main purpose of this paper is to develop the variational structure
of the general Euler-Poincar\'e equations that are the Lagrangian
counterpart to the Lie-Poisson equations associated with a semidirect
product and to show how this is related to the general theory of
Lagrangian reduction. We also want to explain the abstract notion of
circulation that gives rise to a general Kelvin-Noether theorem.

To accomplish these goals, it will be convenient to first recall some
simpler situations, namely the two  general ways of abstracting the
classical Euler equations for a fluid or a rigid body, namely
Lie-Poisson systems on the dual of a Lie algebra and their Lagrangian
counterpart, the ``pure'' Euler-Poincar\'e equations on a Lie algebra.

\paragraph{The Lie-Poisson Equations.}
The Lie-Poisson equations are Hamiltonian equations on the dual of a
Lie algebra and represent an abstraction of the Euler equations for a
rigid body (in body momentum representation) as well as the Euler
equations for an ideal incompressible fluid (in spatial
representation). This set up and its counterpart, the Euler-Poincar\'e
equations on the Lie algebra are the basic ingredients used in the
fundamental paper of Arnold [1966a].

Ignoring, for simplicity, function space technicalities in the infinite
dimensional case (see Ebin and Marsden [1970] for one approach to
dealing with them), we let $G$ be a Lie group with Lie algebra $
\mathfrak{g} $ and let
$F, K$ be real valued functions on the dual space  $\mathfrak{g}
^{\ast}$.  Denoting elements of $\mathfrak{g} ^{\ast}$ by
$\mu$, let the functional derivative of
$F$   at $\mu$ be the unique element $\delta  F/ \delta \mu$ of
$\mathfrak{g}$ defined by
%-----------------------------
\begin{equation}\label{fd}
\lim_{\varepsilon \rightarrow 0}
\frac{1}{\varepsilon }[F(\mu
+ \varepsilon \delta \mu) - F(\mu )]
=  \left\langle \delta  \mu , \frac{
\delta  F}{\delta  \mu } \right\rangle   ,
\end{equation}
%-----------------------------
for all $\delta\mu \in \mathfrak{g}^*$, where $\left\langle \,,
\right\rangle  $ denotes the pairing between
$\mathfrak{g} ^{\ast} $ and  $\mathfrak{g} $. Define the
$(\pm)$  {\bfi Lie-Poisson brackets} by
%-----------------------------
\begin{equation}\label{lpb}
\{F, G\}_\pm (\mu )  =  \pm \left\langle \mu , \left[ \frac{ \delta
F}{\delta  \mu},
\frac{\delta  G}{\delta \mu } \right] \right\rangle   .
\end{equation}
%-----------------------------
These brackets, discovered by Lie [1890], make $\mathfrak{g}^\ast$ into a
Poisson manifold and the Hamiltonian equations associated with a given
Hamiltonian $H$ are called the {\bfi Lie-Poisson equations}. Thus, the
Lie-Poisson equations are determined
by $\dot F = \{ F,H\} $ for all $F$.

We recall these equations in the finite dimensional case in
coordinates. Choose a basis $e_1 , \ldots, e_r$   of
$\mathfrak{g}$ (so ${\rm dim}\,{\mathfrak{g}} = r)$.
Define, as usual, the structure constants
$C^d_{ab}$ of the Lie algebra by
$[e_a, e_b]  =  C^d_{ab}e_d,$
where $a,b$ run from $1$ to $r$ and a sum on $d$ is understood. If $\xi \in
\mathfrak{g}$, its components relative to this basis are denoted $\xi ^a$,
so $ \xi = \xi ^a e _a $.   If $e^1, \ldots, e^n$ is the corresponding dual
basis, and we write
$ \mu = \mu _a e ^a $ (with a sum understood), the
$ (\pm)$ Lie-Poisson brackets become
%-----------------------------
\begin{equation}\label{lpbc}
\{F, K\} _\pm (\mu) =  \pm  C^d_{ab} \mu_d
\frac{\partial F}{\partial
\mu_a}\frac{ \partial K}{\partial \mu_b}.
\end{equation}
%-----------------------------
The Lie-Poisson equations are
$$
\dot \mu _a = \{\mu _a, H\} _\pm (\mu)
= \pm C^d_{ab } \mu _d \frac{\partial H}{\partial \mu _b},
$$
or intrinsically,
%-----------------------------
\begin{equation}\label{alpe }
\dot \mu  = \mp {\rm ad}^{* }_{\delta H/\delta \mu}\, \mu .
\end{equation}
%-----------------------------
Here ${\rm ad}_{\xi} : \mathfrak{g} \to
\mathfrak{g}$ is the adjoint map $\eta \mapsto [\xi, \eta]$,
and ${\rm ad}_{\xi}^{*}: \mathfrak{g}^{*} \to
\mathfrak{g}^{*}$ is its dual.

However, the point of view of general Poisson manifolds for these
systems is sometimes misleading, at least for applications in
mechanics. What is more insightful, and which has its roots in Arnold
[1966a] is the point of view that the {\it Lie-Poisson brackets arise
from canonical brackets on the cotangent bundle $T^\ast G$ by
reduction}. Namely, if we identify $\mathfrak{g}^\ast $ with the
natural Poisson quotient $T^\ast G /G $ using the left action of $G$,
we get the minus Lie-Poisson bracket, while the quotient with the
right action gives the plus Lie-Poisson bracket.

The rigid body is naturally a {\it left} invariant system on $ T^\ast {\rm
SO(3)}$, while ideal fluids naturally give a {\it right} invariant system
on the group of volume preserving diffeomorphisms of the fluid
within the container.

\paragraph{Poincar\'{e} and the Euler equations.} Poincar\'{e}'s work on
the gravitating fluid problem continues the line of investigation begun
by MacLaurin, Jacobi and Riemann and was a natural precursor to his famous
paper, Poincar\'{e} [1901], in which he laid out the basic equations of
Euler type on Lie algebras. He was certainly aware that this
formalism includes the rigid body, heavy top and ideal fluids as special
cases.

To state the Euler-Poincar\'e equations, let $\mathfrak{g}$ be
a given Lie algebra and let $ l : \mathfrak{g} \rightarrow
\mathbb{R} $ be a given function (the La\-gran\-gian), let $\xi$ be a
point in $\mathfrak{g}$ and let $f \in \mathfrak{g}^\ast$ be a
given force (the nature of which we shall explicate later).
Then the Euler-Poincar\'e equations are the following equations
for the evolution of the basic variable $\xi$:
$$
\frac{d}{dt} \frac{\delta l}{\delta  \xi} =
{\rm ad}_{\xi}^{*}\frac{\delta l}{\delta  \xi} + f.
$$
The notation is as follows: $\delta  l / \delta  \xi \in
\mathfrak{g}^\ast$ is the derivative of
$l$ with respect to $\xi$---the dual to the derivative notation used for
the Lie-Poisson equations.

As with the Lie-Poisson setting, these equations are valid for either
finite or infinite dimensional Lie algebras. For fluids, Poincar\'{e} was
aware that one needs to use infinite dimensional Lie algebras, as is clear
from Poincar\'{e} [1910]. The equations also come in two flavors, one for
left invariant systems (the one given) and one for right invariant systems in
which case one changes the sign in front of the adjoint operator (or
changes conventions in the Lie algebra brackets).

In the finite dimensional case, the equations read
\begin{equation}\label{aepe}
\frac{d }{d t} \frac{\partial l}{\partial \xi ^d } =  C ^b _{ a d}
\frac{\partial l}{\partial \xi ^b }  \xi ^a + f _d .
\end{equation}
For example, consider the Lie algebra $\mathbb{R}^3$ with the
usual vector cross product. (Of course, this is the Lie algebra of
the proper rotation group in $ \mathbb{R}^3$.) For $l: \mathbb{R}^3
\rightarrow \mathbb{R}$, the Euler-Poincar\'{e} equations become
\[
  \frac{ d}{dt} \frac{\partial l}{\partial \Omega}
     = \frac{\partial l}{\partial \Omega} \times \Omega + f,
\]
which generalize the Euler equations for rigid body motion to include
external forces and Lagrangians that needn't be quadratic.

These equations were written down for a certain class of
Lagrangians $l$ by Lagrange [1788, Volume 2, Equation A on p.
212], while it  was Poincar\'{e} [1901] who generalized them
(without reference to the ungeometric Lagrange!) to any Lie
algebra. However, La\-gran\-ge did grapple (in much of volume 2 of his
treatise) with the derivation and deeper understanding of the nature of
these equations. While Poincar\'{e} may have understood how to derive
them from other principles, he did not reveal this.

Of course, there was a lot of mechanics going on in the decades
leading up to Poincar\'{e}'s time. It is a curious historical
fact that the Euler-Poincar\'e equations were not pursued extensively and
systematically until rather recently. While many authors mentioned them
(see, e.g., Hamel [1904, 1949] and Chetayev [1941]), it was not until the
Arnold school that a deeper understanding was achieved and was used for
purposes of hydrodynamical stability (see Arnold [1966b, 1988] and
Arnold and Khesin [1992, 1997]).

We now recall the derivation of the ``pure'' Euler--Poincar\'e equations
(i.e., the Euler-Poincar\'e equations with no forcing or advected
terms) for left--invariant Lagrangians on Lie groups (see Marsden
and Scheurle [1993a,b], Marsden and Ratiu [1994], and Bloch et
al. [1996]). The Lagrangian counterpart to the Lie-Poisson reduction of
Poisson manifolds mentioned above is the following:

\begin{thm} Let $G$ be a Lie group and $L : TG \rightarrow \mathbb{R}$ a
left (right) invariant Lagrangian.  Let $l: {\mathfrak{g}}
\rightarrow \mathbb{R}  $ be its restriction to the tangent space at
the identity. For a curve $g(t) \in G,$ let $\xi (t) = g(t) ^{ -1}
\dot{g}(t)$ (respectively $\xi (t) = \dot{g}(t)g(t) ^{ -1}$).  Then
the following are equivalent:
\begin{enumerate}
\item [{\bf i}]
  Hamilton's principle
\begin{equation} \label{eulerpoincare22}
\delta \int _a ^b L(g(t), \dot{g} (t)) dt = 0
\end{equation}  holds, as usual, for variations $\delta g(t)$
of $ g (t) $ vanishing at the endpoints.
\item [{\bf ii}]
      The curve $g(t)$ satisfies the Euler-Lagrange
equations for $L$ on $G$.
\item [{\bf iii}]
      The ``variational'' principle
\begin{equation} \label{eulerpoincare24}
\delta \int _a ^b  l(\xi(t)) dt = 0
\end{equation}  holds on $\mathfrak{g}$, using variations of the
form
\begin{equation} \label{eulerpoincare25}
\delta \xi = \dot{\eta } \pm [\xi , \eta ],
\end{equation}  where $\eta $ vanishes at the endpoints ($+$
corresponds to left invariance and $-$ to right
invariance).\footnote{Because there are constraints on the variations, this
principle is more like a La\-gran\-ge d'Al\-em\-bert principle, which
is why we put ``variational'' in quotes. Of course such problems are
not literally variational.}
\item [{\bf iv}]
        The {\bfi pure Euler-Poincar\'{e} equations\/} hold
\begin{equation} \label{eulerpoincare23}
\frac{ d}{dt} \frac{ \delta l }{ \delta \xi} = \pm
\mbox {\rm ad} _{ \xi} ^{ \ast}
\frac{ \delta l }{ \delta \xi}\,.
\end{equation}
\end{enumerate}
\end{thm}

We make some comments on the proof. First of all, the equivalence of {\bf i}
and {\bf ii} holds on the tangent bundle of any configuration manifold
$Q$, by the general Hamilton principle. To see that {\bf ii} and
{\bf iv} are equivalent, one needs to compute the variations $
\delta \xi $ induced on $\xi = g ^{-1} \dot{ g} = TL _{ g ^{-1}}
\dot{ g} $ by a variation of  $g$.  We will do this for matrix groups; see
Bloch, Krishnaprasad, Marsden, and Ratiu [1994, 1996] for the general case.
To calculate this, we need to differentiate
$ g ^{-1}
\dot{ g}
$ in the direction of a variation
$ \delta g $. If $
\delta g = d g / d \epsilon $ at $ \epsilon = 0 $, where $g$ is
extended to a curve $ g _\epsilon $, then,
\[ \delta \xi = \frac{d }{d \epsilon } g ^{-1} \frac{d }{d t} g, \]
while if $ \eta = g ^{-1} \delta g $, then
\[ \dot{ \eta} = \frac{d }{d t} g ^{-1} \frac{d }{d \epsilon } g . \]
The difference $
\delta \xi - \dot{ \eta} $ is thus the commutator $ [\xi, \eta] $.

To complete the proof, we show the equivalence of {\bf iii} and {\bf
iv}. Indeed, using the definitions and integrating by parts,
%-----------------------------
\begin{align*}
\delta \int l (\xi) d t
    & =  \int \frac{ \delta l }{ \delta \xi}\delta \xi \, d t  =
\int \frac{ \delta l }{ \delta \xi} (\dot{ \eta}
             + {\rm a d} _\xi \eta)\, d t \\
    & =  \int\left[ - \frac{d }{d t} \left( \frac{ \delta l }
        { \delta \xi} \right) +{\rm ad} ^{\ast} _\xi
        \frac{ \delta l }{ \delta \xi} \right] \eta \, dt
\end{align*}
%-----------------------------
so the result follows.

Since the Euler-Lagrange and Hamilton equations on $ TQ$ and $
T^\ast Q$ are equivalent for regular La\-gran\-gians or Hamiltonians, it
follows that the Lie-Poisson and Euler-Poincar\'{e} equations are also
equivalent under a regularity condition. To see this {\em directly\/},
we make the following Legendre transformation from $\mathfrak{g}$ to
$\mathfrak{g} ^{\ast}$:
\[
\mu = \frac{ \delta l }{ \delta \xi}, \quad h (\mu) = \left\langle  \mu, \xi
\right\rangle  - l (\xi) .
\]   Note that
\[
\frac{ \delta h }{ \delta \mu} = \xi + \left\langle \mu, \frac{ \delta
\xi }{ \delta \mu}
\right\rangle  - \left\langle \frac{ \delta l }{ \delta \xi}, \frac{
\delta \xi }{ \delta \mu}
\right\rangle  = \xi
\]
and so it is now clear that the Lie-Poisson and Euler-Poincar\'{e}
equations are equivalent. However, we shall see shortly that this
{\it does not work} for examples like the heavy top, MHD etc., when
we write the equations as Lie-Poisson equations on the dual of a
semidirect product; the reason is that in these examples, the
Hamiltonian is degenerate, i.e., if the matrix of
second derivatives $H_{p_ip_j}$ is singular. This is a fundamental
obstacle and may be viewed as one reason that the La\-gran\-gian
version of the semidirect product theory is interesting. We shall come
to this shortly.

We close this section by mentioning that many other systems can
be put into Euler-Poincar\'e form. For example, this can be done
for the KdV equations as has been shown by Ovsienko and Khesin
[1987] (an account may be found in Marsden and Ratiu [1994]). The
same has been shown for many other shallow water equations and
equations of continuum mechanics, including the equations of
plasma physics (cf. Holm, Marsden and Ratiu [1997] and Cendra, Holm, Hoyle
and Marsden [1997]).

\paragraph{Plan of the paper.} With the above background in hand, we
first set the stage for the main goal of the paper, the comparison of
the general Euler-Poincar\'e equations with Lagrangian reduction, by
describing some of the ingredients in more detail. The first of these
is the Hamiltonian theory of semidirect product reduction in \S2. The
next main ingredient, in \S3, is the derivation of the general
Euler-Poincar\'e equations by the method of reduction of variational
principles. In \S4 we give the Kelvin-Noether theorem; while not
essential to the main goal of the paper, it is a fundamental result
for the Euler-Poincar\'e equations, so it is given for completeness.
In \S5 we describe the general theory of Lagrangian reduction and
illustrate the methods with Wong's equations and use Lagrangian
reduction to give a simple proof of the falling cat theorem of
Montgomery. Finally, in \S6, we show how the Euler-Poincar\'e
equations are linked with the general theory of Lagrangian reduction.

\section{Semidirect Product Reduction}
Before getting to the La\-gran\-gian analogue of semidirect product
reduction, it is useful to summarize the Hamiltonian theory.

The theory begins with the understanding of various examples, such
as the heavy top, ideal compressible fluids and MHD (magnetohydrodynamics).
Building on these examples, the {\it general} study of Lie-Poisson
equations for systems on the dual of a semidirect product Lie algebra was
developed by many authors such as Sudarshan and Mukunda
[1974], Vinogradov and Kupershmidt [1977], Ratiu [1980], Guillemin and
Sternberg [1980], Ratiu [1981, 1982], Marsden [1982], Marsden, Weinstein,
Ratiu, Schmidt and Spencer [1983], Holm and Kupershmidt [1983], Kupershmidt
and Ratiu [1983], Holmes and Marsden [1983], Marsden, Ratiu and Weinstein
[1984a,b], Guillemin and Sternberg [1980, 1984], Holm, Marsden, Ratiu and
Weinstein [1985], Abarbanel, Holm, Marsden, and Ratiu [1986],
Leonard and Marsden [1997] and Marsden, Misiolek, Perlmutter and
Ratiu [1997]. As these and related references show, these equations
apply to a wide variety of systems such as the heavy top,
compressible flow, stratified incompressible flow, MHD and the
dynamics of underwater vehicles.

\paragraph{Generalities on Semidirect Products.} We first recall some
definitions. Let $V$ be a vector space and assume that the Lie group
$G$ acts {\it on the left\/} by linear maps on $V$ (and hence $G$
also acts on on the left on its dual space $V^\ast$). As sets, the
semidirect product $ S = G \,\circledS\, V $ is the Cartesian product
$S  = G \times V$ whose group multiplication is given by
\begin{equation}\label{semidirectleft}
(g_1, v_1) (g_2, v_2) = (g_1 g_2, v_1 + g_1 v_2),
\end{equation}
where the action of $g \in G $ on $v  \in V $ is denoted
simply as $gv$. The identity element is
$ (e,0) $ where $e$ is the identity in $G$ and inversion is
$ (g,v) ^{-1} = ( g ^{-1} , - g ^{-1} v).$

The Lie algebra of $S$ is the semidirect product
Lie algebra,
$\mathfrak{s}    = \mathfrak{g}  \,\circledS\, V $, with the bracket
\begin{equation}\label{semidirectalgebraleft}
[(\xi_1,v _1), (\xi_2, v_2)]
= ([\xi_1,\xi_2],\, \xi_1v_2 - \xi_2 v_1)\,,
\end{equation}
where we denote actions, such as the induced action of $\mathfrak{g}$
on $V$ by concatenation, as in $\xi_1 v_2$.

The adjoint and the coadjoint actions for semidirect products
are given by (see, e.g., Marsden, Ratiu and Weinstein [1984a,b]):
\begin{equation}\label{adjointleft}
(g,v) (\xi, u) = (g \xi, gu - (g \xi) v ).
\end{equation}
and
\begin{equation}\label{coadjointleft}
(g,v) (\mu, a)  = (g \mu  + \rho_v ^\ast(ga), ga),
\end{equation}
where $(g,v)  \in S  = G  \times V $, $( \xi, u)
\in\mathfrak{s} = \mathfrak{g}  \times V$, $(\mu, a)  \in
\mathfrak{s}^\ast = \mathfrak{g}^{\ast}  \times V ^\ast$,
$g\xi = {\rm Ad}_g\xi$, $g\mu =
{\rm Ad}^\ast_{g^{-1}}\mu$, $ga$ denotes the induced
{\it left\/} action of $g$ on $a$ (the {\it left} action of
$G$ on $V$ induces a {\it left} action of $G$ on $ V ^\ast $
--- the inverse of the transpose of the action on $V$),
$\rho _v: \mathfrak{g}  \rightarrow V$ is the linear map
given by $\rho_v (\xi) = \xi v$, and
$\rho _v^\ast: V ^\ast
\rightarrow \mathfrak{g}^{\ast}$ is its dual. For $a \in V ^\ast $,
we shall write, for notational convenience,
\[
   \rho _v^\ast a = v \diamond a  \in \mathfrak{g}^\ast \, ,
\]
which is a bilinear operation in $v$ and $a$. Continuing to employ
the concatenation notation for Lie group or algebra actions, the
identity
\begin{equation}\label{diamond-def}
\left \langle \eta a, v
\right\rangle  = - \left \langle v \diamond a\,,
\eta \right\rangle
\end{equation}
for all $ v \in V$, $ a \in V^\ast $ and
$ \eta \in \mathfrak{g} $ is another way to write the definition of
$ v \diamond a \in \mathfrak{g}^\ast$. Using this
notation, the coadjoint action reads
$   (g,v) (\mu, a)  = (g \mu  + v \diamond (ga), ga).$

When working with various models of continuum mechanics and plasmas
it is convenient to work with {\it right\/} representations of $G$
on the vector space $V$ (as in, for example, Holm, Marsden and Ratiu
[1986]). In this context of course the above formalism must be
suitably modified.

\paragraph{Lie-Poisson Brackets and Hamiltonian Vector Fields.}
For a {\it left\/} representation of $G$ on $V$ the
$\pm$ Lie-Poisson bracket of two functions $f, k : \mathfrak s^\ast
\rightarrow \mathbb{R} $ is given by
\begin{eqnarray}\label{leftLP}
\{f, k\} _\pm (\mu, a) & = & \pm \left\langle \mu,
\left[
\frac{\delta  f}{\delta \mu } , \frac{\delta k}{\delta \mu }
\right] \right\rangle
\pm
\left\langle a, \frac{\delta f}{\delta \mu } \frac{\delta
k }{ \delta a} - \frac{\delta
k}{\delta \mu } \frac{\delta f}{\delta a} \right\rangle
\end{eqnarray}
where $\delta f / \delta \mu  \in \mathfrak{g}$, and
$\delta f / \delta a \in V$
are the functional derivatives of $f$.
The Hamiltonian vector field of $h :
\mathfrak s^\ast \rightarrow \mathbb{R}$ has the expression
\begin{equation}\label{leftham}
X_h (\mu, a) = \mp \left( {\rm ad}^\ast _{\delta h /\delta \mu}\mu
-\frac{\delta h }{ \delta a} \diamond  a,\,
-\,\frac{ \delta h}{\delta \mu}\, a \right)
\end{equation}
Thus, Hamilton's equations (the Lie-Poisson equations) on the dual of
a semidirect product are given by
\begin{eqnarray} \label{leftsemi1.eqn}
\dot{ \mu } & = &  \mp\,  {\rm ad}^\ast _{\delta h /\delta \mu}\mu
\pm  \frac{\delta h }{ \delta a} \diamond a\,, \\
\dot{ a } & =& \pm\, \frac{\delta h}{\delta \mu}\, a ,
\label{leftsemi2.eqn}
\end{eqnarray}
where overdot denotes time derivative.
Again, for {\it right\/} representations of $G$ on $V$ the above
formulae must be appropriately modified.

\paragraph{Symplectic Actions by Semidirect Products.}
We consider a (left) symplectic action of
$S$ on a symplectic manifold $P$ and assume that this action has
an equivariant momentum map ${\bf J}_S : P \rightarrow\mathfrak{s}
^{\ast}$.  Since $V$ is a (normal) subgroup of $S$, it also acts
on $P$ and has a momentum map
${\bf J}_V: P  \rightarrow V ^\ast$ given by
$ {\bf J}_V  = i _V ^\ast \circ {\bf J}_S\,,$ where $i _V : V
\rightarrow \mathfrak{s}  $ is the inclusion
$v \mapsto (0,v)$ and $i _V ^\ast: \mathfrak{s}  ^\ast  \rightarrow
V ^\ast$ is its dual. We think of this merely as saying that
${\bf J}_V$ is the second component of ${\bf J}_S$.

We can regard $G$ as a subgroup of $S$ by $g \mapsto (g,0)$.
Thus, $G$ also has a momentum map that is the first component
of ${\bf J}_S$ but this will play a secondary role in what
follows. On the other hand, equivariance of ${\bf J}_S$ under
$G$ implies the following relation for ${\bf J}_V$:
\begin{equation} \label{relation.eq}
{\bf J}_V(gz)  = g {\bf J}_V(z)
\end{equation}
where we denote the appropriate action of  $g \in G$ on an
element by concatenation, as before. To prove (\ref{relation.eq}),
one uses the fact that for the coadjoint action of $S$ on
$\mathfrak{s} ^\ast$ the second component is just the dual of
the given action of $G$ on $V$.

\paragraph{The Classical Semidirect Product Reduction Theorem.}
In a number of interesting applications such as compressible fluids,
the heavy top, MHD, etc., one has two symmetry groups that do not
commute and commuting reduction by stages (a theorem from Marsden
and Weinstein [1974]) does not apply. In this more general
situation, it matters in what order one performs the reduction,
which occurs, in particular for semidirect products. The main result
covering the case of semidirect products has a complicated history,
some of which has been sketched; we follow the version of
Marsden, Ratiu and Weinstein [1984a,b].

The semidirect product reduction theorem states, roughly speaking,
that for the semidirect product $S = G\,\circledS\, V$ where $G$ is a
group acting on a vector space $V$ and $S$ is the semidirect product,
one can first reduce $T^{\ast}S$  by $V$ and then by $G$ and obtain
the same result as reducing by $S$. As above, we let $\mathfrak{s} =
\mathfrak{g}\,\circledS\, V$ denote the Lie algebra of $S$. The precise
statement is as follows.

\begin{thm}[Semidirect Product Reduction Theorem.]
\label{semidirect.thm} \quad Let $S = G \,\circledS\, V$,
choose $\sigma = (\mu, a) \in \mathfrak{g}^{\ast} \times  V
^{\ast}$, and reduce $T ^{\ast}S $ by the action of $S$ at $\sigma$
giving the coadjoint orbit $ {\mathcal O}_\sigma $ through
$\sigma\in
\mathfrak{s}^\ast$.  There is a symplectic diffeomorphism between
${\mathcal O}_\sigma $ and the reduced space obtained by reducing
$T^{\ast}G$ by the subgroup $G_a$ (the isotropy of $G$ for its
action on $V ^\ast$ at the point $a\in V ^{\ast}$) at the point
$\mu_a \in \mathfrak{g}_a^\ast$ defined by restriction:
$\mu_a = \mu |\mathfrak{g}_a$, where $\mathfrak{g}_a$ is the Lie
algebra of $G_a$.
\end{thm}

\paragraph{Reduction by Stages.} The preceding result is a special
case of a general theorem on reduction by stages for semidirect
products acting on a symplectic manifold (See Marsden, Misiolek,
Perlmutter and Ratiu [1997] for this and more general results
dealing with {\it group extensions} and see Leonard and Marsden
[1997] for an application to underwater vehicle dynamics.)

As above, consider a symplectic action of $S$ on a symplectic
manifold $P$ and assume that this action has an equivariant momentum
map ${\bf J}_S : P \rightarrow\mathfrak{s}^{\ast}$. As we have
explained, the momentum map for the action of
$V$ is the map ${\bf J}_V: P  \rightarrow V ^\ast$ given by
${\bf J}_V  = i _V ^\ast \circ {\bf J}_S$

We carry out the symplectic reduction of $P$ by $S$ at
a regular value $\sigma=(\mu, a)$ of the momentum map ${\bf J}_S$
for $S$ in two stages. First, reduce
$P$ by $V$ at the value $a$ (assume it to be a regular value) to
get the reduced space $P_a := {\bf J}_V^{-1} (a)/V$. Second, form
the group $G_a$ consisting of elements of $G$ that leave the point
$a$ fixed using the action of $G$ on $V^\ast$. One shows (and this
step is not trivial) that the group $G_a$ acts on $P_a$ and has an
induced equivariant momentum map ${\bf J}_a : P _a \rightarrow
\mathfrak{g}^{\ast} _a$, where $\mathfrak{g}_a$ is the Lie algebra
of $G_a$, so one can reduce
$P_a$ at the point $\mu_a : = \mu | \mathfrak{g} _a$ to get the
reduced space $(P_a)_{\mu_a}  = {\bf J}_a^{-1}(\mu_a) /
(G_a)_{\mu_a}$.

\begin{thm}[Reduction by Stages for Semidirect Products.] The
reduced space $(P_a)_{\mu_a}$ is symplectically
diffeomorphic to the reduced space $P_\sigma$ obtained by
reducing $P$ by $S$ at the point $\sigma = (\mu,a)$.
\end{thm}

\paragraph{Semidirect Product Reduction of Dynamics.} There is a
technique for reducing dynamics that is associated with the geometry
of the semidirect product reduction theorem. We start with a
Hamiltonian $H_{a_0}$  on $T^{\ast}G$ that depends parametrically on
a variable $a _0 \in V^{\ast}$. The Hamiltonian, regarded as a map
$H : T^{\ast}G \times V ^{\ast} \rightarrow \mathbb{R}$
is assumed to be invariant on $T^{\ast}G$ under the action of $G$
on $T^{\ast}G\times V^{\ast}$.
This condition is equivalent to the
invariance of the function $H$ defined on
$ T^{\ast} S = T ^{\ast} G \times V \times V^{\ast}$
extended to be constant in the variable $V$ under the action of the
semidirect product. By the semidirect product reduction theorem, the
dynamics of $ H _{a_0} $ reduced by $ G_{a_0} $, the isotropy group
of $a_0$, is symplectically equivalent to Lie-Poisson dynamics on
$\mathfrak{s}^\ast = \mathfrak{g}^{\ast}\times V^\ast$. This
Lie-Poisson dynamics is given by the equations (\ref{leftsemi1.eqn})
and (\ref{leftsemi2.eqn}) for the function
$ h( \mu, a ) = H (\alpha _g , g^{-1} a) $ where
$ \mu = g ^{-1} \alpha _g $.

\paragraph{Cotangent bundle reduction.} It will be useful to recall
a few additional facts about cotangent bundle reduction.
Following standard notation consistent with what we have already
used, a symplectic reduced space at momentum value $\mu$ is
denoted $ P _\mu := {\bf J} ^{-1} (\mu) / G _\mu $. For cotangent
bundle reduction we are considering $ P = T^\ast Q $ and the action
of a Lie group $G$ on $Q$ with the standard momentum map. The
simplest case of cotangent bundle reduction is reduction at the
momentum value zero in which case one has $(T ^\ast Q )_{\mu = 0} =
T ^\ast (Q/G)$, the latter with the canonical symplectic form. If
$G$ is abelian, then $(T ^\ast Q )_\mu \cong T ^\ast (Q/G )$ but the
latter has a symplectic structure modified by magnetic terms; that
is, by the curvature of the mechanical connection. This abelian
version of cotangent bundle reduction was developed by Smale [1970]
and Satzer [1977] and was generalized to the nonabelian case in
Abraham and Marsden [1978]. Kummer [1981] introduced the
interpretations of these results in terms of the mechanical
connection. This set up was effectively used, for example, in
Guichardet [1984] and Iwai [1987] to study geometric phases in classical
molecular dynamics. For additional information on the cotangent bundle
reduction theorem, see Marsden [1992].

When combined with the cotangent bundle reduction theorem,
the semidirect product reduction theorem is a useful tool. For
example this shows directly that the generic coadjoint orbits for the
Euclidean group are cotangent bundles of spheres with the associated
coadjoint orbit symplectic structure given by the canonical
structure plus a magnetic term. In fact, this technique allows one
to understand the analogue of this geometrical structure for general
semidirect products.

The ``bundle picture'' begun by these early works was significantly
developed by Montgomery, Marsden and Ratiu [1984] and by Montgomery
[1986], and was motivated by work of Weinstein and Sternberg on
Wong's equations (the equations for a particle moving in a Yang-Mills
field). We shall come to the La\-gran\-gian counterpart of this
theory shortly.

\section{Lagrangian Semidirect Product Theory}
\paragraph{Introduction.} Despite all the activity on the
Hamiltonian theory of semidirect products, little attention was paid
to the corresponding Lagrangian side. Now that Lagrangian reduction
is maturing (see Marsden and Scheurle [1993a,b]) and useful
applications are emerging such as to Hamilton's principle asymptotics (see
Holm [1996]) and to numerical algorithms (Wendlandt and Marsden
[1997] and Marsden, Patrick and Shkoller [1997]), it is appropriate
to consider the corresponding Lagrangian question more deeply.

The theory is entirely based on variational
principles with symmetry. Consistent with the variational
formulation, note that none of the theorems in this section require
that the Lagrangian be nondegenerate. The theory is not dependent on
the previous Hamiltonian formulation, although we shall, of course,
make links with it under the appropriate regularity assumptions.

The theorems that follow are modeled after the reduction theorem
for the {\it pure} Euler-Poincar\'e equations given in the
introduction, although, as we shall explain, they are {\it not\/}
literally special cases of it. As in the Hamiltonian case, the main
distinction between the pure Euler-Poincar\'e equations and the
general ones is the presence of advected quantities, such as the
body representation of the direction of gravity in the heavy top and
the density in compressible fluids.

Abstractly, these advected quantities are reflected by the fact that
the Lagrangian $L$ and its reduction $l$, depend on another parameter
$a\in V^\ast$, where, as in the Hamiltonian case, $V$ is a
representation space for the Lie group $G$ and $L$ has an invariance
property relative to both arguments.

As we shall see shortly, the resulting Euler--Poincar\'e
equations are {\it not\/} the same as the pure Euler--Poincar\'e
equations for the semidirect product Lie algebra
$\mathfrak{g}\,\circledS\, V^\ast$.

\paragraph{The basic ingredients.}
We begin with  a {\it left\/} representation of Lie group $G$ on
the vector space $V$ and $G$ acts in the natural way on the {\it
left\/} on $TG \times V^\ast$: $h(v_g, a) = (hv_g, ha)$.
We assume that we have a  $ L : T G \times V ^\ast
\rightarrow \mathbb{R}$ is left $G$--invariant.
In particular, if $a_0 \in V^\ast$, define the
Lagrangian $L_{a_0} : TG \rightarrow \mathbb{R}$ by
$L_{a_0}(v_g) = L(v_g, a_0)$. Then $L_{a_0}$ is left
invariant under the lift to $TG$ of the left action of
$G_{a_0}$ on $G$, where $G_{a_0}$ is the
isotropy group of $a_0$.

Left $G$--invariance of $L$ permits us to define the reduced
La\-gran\-gian
$l: {\mathfrak{g}} \times V^\ast \rightarrow \mathbb{R}$ by
$l(g^{-1} v_g, g^{-1} a) = L(v_g, a). $
Conversely,  this relation defines for any
$l: {\mathfrak{g}} \times V^\ast \rightarrow
\mathbb{R} $ a left $G$--invariant function
$ L : T G \times V ^\ast
\rightarrow \mathbb{R} $.

For a curve $g(t) \in G, $ let
$ \xi (t) := g(t) ^{ -1} \dot{g}(t)$ and define the curve
$a(t)$ as the unique solution of the following linear
differential equation with time dependent coefficients
$ \dot a(t) = -\xi(t) a(t),$
with initial condition $a(0) = a_0$. The solution can be
written as $a(t) = g(t)^{-1}a_0$.

\paragraph{The Euler-Poincar\'e equations.} The generalization of
the pure Euler-Poincar\'e theorem is the following.
\begin{thm} \label{lall}
With the preceding notation, the following are equivalent:
\begin{enumerate}
\item [{\bf i} ] With $a_0$ held fixed, Hamilton's variational
principle
\begin{equation} \label{hamiltonprinciple}
\delta \int _{t_1} ^{t_2} L_{a_0}(g(t), \dot{g} (t)) dt = 0
\end{equation}
holds, for variations $\delta g(t)$
of $ g (t) $ vanishing at the endpoints.
\item [{\bf ii}  ] $g(t)$ satisfies the Euler--Lagrange
equations for $L_{a_0}$ on $G$.
\item [{\bf iii} ]  The constrained variational principle
(of La\-gran\-ge d'Al\-em\-bert type)
\begin{equation} \label{variationalprinciple}
\delta \int _{t_1} ^{t_2}  l(\xi(t), a(t)) dt = 0
\end{equation}
holds on $\mathfrak{g} \times V ^\ast $, using variations of $ \xi $
and
$a$ of the form
\begin{equation} \label{epvariations}
\delta \xi = \dot{\eta } + [\xi , \eta ], \quad
\delta a =  -\eta a ,
\end{equation}
where $\eta(t) \in \mathfrak{g}$ vanishes at the endpoints.
\item [{\bf iv}] The {\bfi Euler--Poincar\'{e}}
equations hold on $\mathfrak{g} \times V^\ast$
\begin{equation} \label{eulerpoincare}
\frac{d}{dt} \frac{\delta l}{\delta \xi} =
\mbox {\rm ad}_{\xi}^{\ast} \frac{ \delta l }{ \delta \xi}
+ \frac{\delta l}{\delta a} \diamond a\,,
\quad \hbox{where}\quad \dot a(t) = -\xi(t) a(t).
\end{equation}
\end{enumerate}
\end{thm}
\vspace{0.2in}

\noindent {\bf Proof.\,} \quad The equivalence of {\bf i} and {\bf ii}
holds for any configuration manifold and so, in particular, it
holds in this case.

Next we show the equivalence of {\bf iii} and {\bf iv}.
Indeed, using the definitions,
integrating by parts, and taking into
account that $\eta(t_1) = \eta (t_2) = 0$, we compute the
variation of the integral to be
\begin{eqnarray*}
\delta \int_{t_1}^{t_2} l(\xi(t), a(t)) dt &=&
\int_{t_1}^{t_2} \left (\left \langle
\frac{ \delta l}{\delta \xi}\,,
\delta \xi \right\rangle  + \left \langle \delta a,
\frac{\delta l}{\delta a}\right\rangle  \right ) \, dt\\  &=&
\int_{t_1}^{t_2} \left ( \left \langle
\frac{\delta l}{\delta \xi}\,, \dot{\eta} +
\mbox {\rm ad}_{\xi} \eta \right\rangle  -
\left \langle \eta a, \frac{\delta l}{\delta a}
\right\rangle  \right )dt\\ &=&
\int_{t_1}^{t_2} \left (\left \langle - \frac{ d}{dt}
\left(\frac{\delta l }{ \delta \xi} \right) +
\mbox {\rm ad}_{\xi}^{\ast}\frac{\delta l}{\delta \xi}\,, \eta
\right\rangle  + \left \langle \frac{\delta l }{\delta a}
\diamond a\,,
\eta \right\rangle  \right )\,dt \\ &=&
\int_{t_1}^{t_2} \left \langle - \frac{ d}{dt}
\left(\frac{ \delta l }{ \delta \xi} \right) +
\mbox {\rm ad}_{\xi}^{\ast}\frac{\delta l}{\delta \xi} +
\frac{\delta l }{\delta a} \diamond a\,, \eta
\right\rangle  \,dt
 \end{eqnarray*}
and so the result follows.

Finally we show that {\bf i} and {\bf iii} are equivalent.
First note that the $G$--invariance of $L:TG
\times V^\ast \rightarrow \mathbb{R}$ and the definition of $a(t) =
g(t)^{-1}a_0$ imply that the
integrands in (\ref{hamiltonprinciple}) and
(\ref{variationalprinciple}) are equal. However, all variations
$\delta g(t) \in TG$ of $g(t)$ with fixed endpoints induce and are
induced by variations $\delta \xi(t) \in
\mathfrak{g}$ of $\xi(t)$ of the form $\delta \xi = \dot{\eta } +
[\xi , \eta ]$ with $\eta(t) \in \mathfrak{g}$ vanishing at the
endpoints; the relation between $\delta g(t)$ and $\eta(t)$
is given by $\eta(t) = g(t)^{-1} \delta g(t)$. This is the content
of the following lemma (which is elementary for matrix groups) proved
in Bloch et al. [1996].

\begin{lem} \label{propvariations} Let $g: U \subset\mathbb{R} ^2
\rightarrow G$ be a smooth map
and denote its partial derivatives by
$ \xi (t,\varepsilon ) = T L _{ g(t, \varepsilon ) ^{ -1}}
(\partial g(t,\varepsilon) / \partial t)$ and
$ \eta (t, \varepsilon) =
TL_{g(t,\varepsilon)^{-1}}(\partial g(t,\varepsilon) /
\partial \varepsilon).$
Then
\begin{equation} \label{eulerpoincare20}
\frac{ \partial \xi }{ \partial \varepsilon} -
\frac{ \partial \eta }{
\partial t} = [\xi , \eta ]\,.
\end{equation}
Conversely, if $U$ is simply connected and $\xi , \eta : U
\rightarrow { \mathfrak{g} }$ are smooth functions
satisfying (\ref{eulerpoincare20})
then there exists a smooth function
$g: U \rightarrow G$ such that $\xi (t,
\varepsilon ) = T L _{ g(t, \varepsilon ) ^{ -1}}
(\partial g(t, \varepsilon ) /
\partial t) $ and $\eta (t,\varepsilon )
= TL _{ g(t,\varepsilon ) ^{ -1}}
(\partial g(t,\varepsilon ) /
\partial \varepsilon ).$
\end{lem}

Thus, if {\bf i} holds, we define $\eta(t) = g(t)^{-1} \delta
g(t)$ for a variation $\delta g(t)$ with fixed endpoints.
Then if we let $\delta \xi = g(t)^{-1} \dot g(t)$, we have by the
above proposition
$\delta \xi = \dot{\eta } + [\xi ,\eta]$.  In addition, the
variation of $a(t) = g(t)^{-1} a_0$ is
$\delta a(t) = -\eta(t) a(t)$. Conversely, if  $\delta \xi =
\dot{\eta } + [\xi ,\eta]$ with $\eta(t)$ vanishing at the
endpoints, we define $\delta g(t) = g(t) \eta(t)$ and the above
proposition guarantees then that this $\delta g(t)$ is the general
variation of $g(t)$ vanishing at the endpoints. From $\delta a(t)
 = -\eta(t)a(t)$ it follows that the variation of $g(t)a(t) = a_0$
vanishes, which is consistent with the dependence of $L_{a_0}$
only on $g(t), \dot g(t)$.
\quad $\blacksquare$

\paragraph{Remark.} The Euler-Poincar\'e equations are not the
{\it pure} Euler-Poincar\'e equations because we are not regarding
$\mathfrak{g}
\times V ^\ast$ as a Lie algebra. Rather these
equations are thought of as a generalization of the classical
Euler-Poisson equations for a heavy top, but written in body
angular velocity variables. Some authors may thus prefer the
term Euler-Poisson-Poincar\'{e} equations for these
equations. The following argument shows that these
Euler--Poincar\'e equations (\ref{eulerpoincare}) are not the pure
Euler--Poincar\'e equations for the semidirect product Lie  algebra
$\mathfrak{g}\,\circledS\, V^\ast$.
Indeed, by (\ref{eulerpoincare23}) the pure Euler--Poincar\'e
equations
\[
\frac{d}{dt}\frac{\delta l}{\delta (\xi, a)} = \mbox {\rm ad}_{(\xi,
a)}^{\ast} \frac{ \delta l }{ \delta (\xi,a)}\,, \quad (\xi, a) \in
\mathfrak{g}\,\circledS\, V^\ast
\]
for $l:\mathfrak{g}\,\circledS\, V^\ast \rightarrow \mathbb{R} $ become
\[
\frac{d}{dt} \frac{\delta l}{\delta \xi} =
\mbox {\rm ad}_{\xi}^{\ast} \frac{ \delta l }{ \delta \xi}
+ \frac{\delta l}{\delta a} \diamond a, \quad
\frac{d}{dt}\frac{\delta l}{\delta a} = -\xi\frac{\delta l}{\delta a}\,,
\]
which is a different system from that given by the
Euler--Poincar\'e equation (\ref{eulerpoincare}) and
$\dot a = -\xi a$, even though the first equations of both systems are
identical.

\paragraph{The Legendre Transformation.} As we explained earlier,
one normally thinks of passing from Euler--Poincar\'e equations
on a Lie algebra
$\mathfrak{g}$ to Lie--Poisson
equations on the dual $\mathfrak{g}^\ast$
by means of the Legendre
transformation. In our case, we start with a
Lagrangian on $ \mathfrak{g}
\times V ^\ast $ and perform a {\it partial}
Legendre transformation in the variable
$ \xi $ only, by writing
\begin{equation}\label{legendre}
\mu = \frac{\delta l}{\delta \xi}\,, \quad
h(\mu, a) = \left\langle  \mu, \xi\right\rangle  - l(\xi, a).
\end{equation}
A simple calculation shows that
$ \delta h / \delta \mu = \xi$, and $\delta h / \delta a = -\delta l
/ \delta a$, so that (\ref{eulerpoincare}) and $\dot a(t) =
-\xi(t) a(t)$ imply (\ref {leftham}) for the {\it minus\/}
Lie--Poisson bracket (that is, the sign + in (\ref {leftham})). If
this Legendre transformation is invertible, then we can also pass
from the minus Lie--Poisson equations (\ref {leftham}) to the
Euler--Poincar\'e equations (\ref{eulerpoincare}) together with the
equations $\dot a(t) = -\xi(t) a(t)$.

\section{The Kelvin-Noether Theorem} In this section, we derive a
version of the Noether theorem that holds for solutions of the
Euler-Poincar\'e equations. Our formulation is motivated and designed
for continuum theories (and hence the name Kelvin-Noether), but it
holds in a more general situation. Of course it is well known (going
back at least to the work of Arnold [1966a]) that the
Kelvin circulation theorem for ideal flow is closely related to the
Noether theorem applied to continua using the particle relabeling
symmetry group.

\paragraph{The Kelvin-Noether Quantity.} Start with a Lagrangian
$L _{a _0}$ depending on a parameter $a _0 \in V ^\ast$ as above.
Consider a manifold ${\mathcal C}$ on which $G$ acts (we assume this
is also a left action) and suppose we have an equivariant map
${\mathcal K} : {\mathcal C} \times V ^\ast
\rightarrow \mathfrak{g} ^{\ast \ast} $.

In the case of continuum theories, the space ${\mathcal
C}$ may be chosen to be a loop space and $\left\langle {\mathcal K}
(c, a), \mu \right\rangle  $ for $c \in {\mathcal C}$ and $\mu \in
\mathfrak{g}^\ast$ will be  a circulation. This class of examples
also shows why we {\it do not} want to identify the double dual
$\mathfrak{g} ^{\ast \ast}$ with $\mathfrak{g}$.

Define the {\bfi Kelvin-Noether quantity}
$I : {\mathcal C} \times \mathfrak{g} \times V ^\ast
\rightarrow \mathbb{R}$ by
\begin{equation}\label{KelvinNoether}
I(c, \xi, a) = \left\langle{\mathcal K} (c, a), \frac{ \delta
l}{\delta \xi}( \xi , a)
\right\rangle  .
\end{equation}

\begin{thm}[Kelvin-Noether] \label{KelvinNoetherthm}Fixing $c_0 \in
{\mathcal C}$, let $\xi (t), a(t)$ satisfy the
Euler-Poincar\'e equations and define $g(t)$ to be the solution of
$\dot{g}(t) = g(t) \xi(t)$ and, say, $ g (0) = e$. Let
$c(t) = g(t)^{-1} c_0$ and $I(t) = I(c(t), \xi(t), a(t))$. Then
\begin{equation}
\frac{d}{dt} I(t) = \left\langle {\mathcal K}(c(t), a (t) ),
                     \frac{\delta l}{\delta a} \diamond a
\right\rangle  .
\end{equation}
\end{thm}

\noindent{\bf Proof.\,} First of all, write $ a (t) = g(t) ^{-1} a _0$
as we did
previously and use equivariance to write $I(t)$ as follows:
\[
   \left\langle {\mathcal K} (c(t), a(t) ) ,
          \frac{\delta l}{\delta \xi} (\xi(t), a(t)) \right\rangle
   =  \left\langle {\mathcal K} ( c_0, a _0 ), g(t)
         \left[ \frac{\delta l}{\delta \xi} (\xi(t),a(t)) \right]
\right\rangle
\]
The $g ^{-1}$ pulls over to the right side as $g$ (and not $g^{-1}$)
because of our conventions of always using left representations.  We
now differentiate the right hand side of this equation. To do so, we
use the following well known formula for differentiating the
coadjoint action (see Marsden and Ratiu [1994], page 276):
\[
\frac{d}{dt} [g(t) \mu (t)]
=  g (t) \left[ - {\rm ad}_{\xi (t)}^\ast  \mu (t)
     +   \frac{d}{dt} \mu (t) \right] ,
\]
where, as usual,
\[ \xi (t) = g (t) ^{-1} \dot{g} (t) .
\]
Using this and the
Euler-Poincar\'e equations, we get
\begin{eqnarray*}
      \frac{d}{dt} I
  & = &  \frac{d}{dt} \left\langle {\mathcal K} ( c_0, a_0 ), g(t)
           \left[ \frac{\delta l}{\delta \xi} (\xi(t),a(t)) \right]
\right\rangle   \\
  & = & \left\langle {\mathcal K} ( c _0, a_0 ) , \frac{d}{dt}
\left\{ g(t)
           \left[ \frac{\delta l}{\delta \xi} (\xi(t),a(t)) \right]
\right\} \right\rangle   \\
  & = & \left\langle
      {\mathcal K} (c_0, a_0), g(t) \left[ - {\rm ad}^\ast _\xi
       \frac{\delta l}{\delta \xi}
      + {\rm ad} _\xi ^\ast \frac{\delta l}{\delta \xi}
      + \frac{\delta l}{\delta a} \diamond a \right]
\right\rangle   \\
& = & \left\langle {\mathcal K} (c_0, a_0 ) , g(t)  \left[
\frac{\delta l}{\delta a} \diamond a\right] \right\rangle   \\
& = & \left\langle {g(t) ^{-1} \mathcal K} ( c _0, a_0 ) , \left[
\frac{\delta l}{\delta a} \diamond a\right] \right\rangle   \\
& = & \left\langle {\mathcal K} ( c (t), a (t)  ) ,  \left[
\frac{\delta l}{\delta a} \diamond a\right] \right\rangle
\end{eqnarray*}
where, in the last steps we used the definitions of the coadjoint action,
the Euler-Poincar\'e equation (\ref{eulerpoincare})
and the equivariance of the map $ {\mathcal K} $.
\quad
$\blacksquare$
\medskip

\begin{cor} For the pure Euler-Poincar\'e equations, the Kelvin quantity
$I (t) $, defined the same way as above but with
$ I : {\mathcal C} \times \mathfrak{g} \rightarrow\mathbb{R}$,
is conserved.
\end{cor}

\paragraph{The Heavy Top.} As a simple illustration of the results
so far, we consider the heavy top. For continuum examples, we refer
to Holm, Marsden and Ratiu [1997] for fluid theories and to Cendra,
Holm, Hoyle and Marsden [1997] for Vlasov plasmas.

The heavy top kinetic energy is given
by the left invariant metric on $SO(3)$ whose value at the identity is
$\left\langle \mathbf{x}, \mathbf{y} \right\rangle
= \mathbb{I}\mathbf{x} \cdot \mathbf{y}$, where $\mathbf{x},
\mathbf{y} \in
\mathbb{R}^3$ are thought of as elements of $\mathfrak{so}(3)$, the
Lie algebra of $SO(3)$, via the isomorphism $\mathbf{x} \in
\mathbb{R}^3 \mapsto \hat{\mathbf{x}} \in \mathfrak {so}(3)$,
${\hat \mathbf{x}}\mathbf{y}:=\mathbf{x}\times \mathbf{y}$, and where
$\mathbb{I}$ is the (time independent) moment of inertia tensor in body
coordinates, usually taken as a diagonal matrix by choosing the body
coordinate system to be a principal axes body frame. This kinetic
energy is thus left invariant under the full group $SO(3)$. The
potential energy is given by the work done in lifting the weight of
the body to the height of its center of mass, with the direction of
gravity pointing downwards. If
$m$ denotes the total mass of the top, $g$ the magnitude of the
gravitational acceleration, $\chi$ the unit vector of the oriented line
segment pointing from the fixed point about which the top
rotates (the origin of a spatial coordinate system) to the
center of mass of the body, and $l$ its length,
then the potential energy is given by $-mgl \mathbf{R}^{-1}\mathbf{e}_3
\cdot \boldsymbol{\chi}$, where $\mathbf{e}_3$ is the axis of the
spatial coordinate system parallel to the direction of gravity
but pointing upwards and where $\mathbf{R} \in {\rm SO(3)}$ is the
orientation of the body.  This potential energy breaks the full
$SO(3)$ symmetry and is invariant only under the rotations
$S^1$ about the $\mathbf{e}_3$--axis.

To apply Theorem \ref{lall}, we take the Lagrangian of the heavy
top to be the kinetic minus the potential energy, regarded as a
function on $TSO(3) \times \mathbb{R}^3 \rightarrow \mathbb{R}$. In
particular, the potential is given by
$
U(u_\mathbf{R}, \mathbf{v}) = mg l \mathbf{R}^{-1} \mathbf{v}  \cdot
\boldsymbol{\chi},
$
where $u_\mathbf{R} \in T _\mathbf{R} {\rm SO(3)}$ (eventually
identified with $\dot{\mathbf{R}}$) and
$\mathbf{v}\in\mathbb{R}^3$ (eventually identified with the direction
of gravity), which is easily seen to be
$SO(3)$--invariant. Thus, the heavy top equations of motion in the
body representation are given by the Euler--Poincar\'e equations
(\ref {eulerpoincare}) for the Lagrangian $l:
\mathfrak {so}(3) \times \mathbb{R}^3 \rightarrow
\mathbb{R} $. To compute the explicit expression of
$l$, denote by $\boldsymbol{\Omega}$ the angular velocity and by
$\boldsymbol{\Pi} =
\mathbb{I} \boldsymbol{\Omega}$ the angular momentum in the body
representation. Let
$\boldsymbol{\Gamma} = \mathbf{R}^{-1}\mathbf{v}$; if $\mathbf{v} =
\mathbf{e}_3$, the unit vector pointing upwards on the vertical spatial
axis, then
$\boldsymbol{\Gamma}$ is this unit vector viewed by an observer fixed
and moving with the body. The Lagrangian $l:\mathfrak{so}(3)
\times \mathbb{R}^3 \rightarrow \mathbb{R}$ is thus given by
\[
l( \boldsymbol{\Omega}, \boldsymbol{\Gamma})
=  L(\mathbf{R}^{-1}u_\mathbf{R}, \mathbf{R}^{-1} \mathbf{v})
= \frac{1}{2} \boldsymbol{\Pi} \cdot \boldsymbol{\Omega} + mgl
\boldsymbol{\Gamma}
\cdot \boldsymbol{\chi}\,.
\]
It is now straightforward to compute the Euler-Poincar\'e equations.
First note that
\[
\frac{\delta l}{\delta \boldsymbol{\Omega}} = \boldsymbol{\Pi}, \quad
\frac{\delta l}{\delta \boldsymbol{\Gamma}} = mgl
\boldsymbol{\chi}\,.
\]
Since
\[
\operatorname{ad}_{\boldsymbol{\Omega}}^{\ast} \boldsymbol{\Pi}
= \boldsymbol{\Pi} \times \boldsymbol{\Omega}\,, \quad \mathbf{v}
\diamond
\boldsymbol{\Gamma} = -\boldsymbol{\Gamma} \times \mathbf{v}\,,
\; \mbox{and} \quad \hat{\Omega}\boldsymbol{\Gamma}
= -\boldsymbol{\Gamma} \times \Omega\,,
\]
the Euler--Poincar\'e equations are
\[
\dot{\boldsymbol{\Pi}} = \boldsymbol{\Pi} \times \boldsymbol{\Omega} +
mgl\boldsymbol{\Gamma}
\times \boldsymbol{\chi}
\]
which are coupled to the $\boldsymbol{\Gamma}$ evolution
$
\dot{\boldsymbol{\Gamma}} = \boldsymbol{\Gamma} \times
\boldsymbol{\Omega}\,.
$
This system of two vector equations for $\Pi,\Gamma$ are the classical
Euler--Poisson equations and they describe the motion of the
heavy top in the body representation.

To illustrate the Kelvin-Noether theorem, choose
${\mathcal C} = \mathfrak{g}$ and let
$ {\mathcal K}: {\mathcal C} \times V ^\ast  \rightarrow
\mathfrak{g}^{\ast
\ast}
\cong \mathfrak{g}$ be the map $ (W, \boldsymbol{\Gamma} )\mapsto W$.
Then the Kelvin-Noether theorem gives the statement
\[
\frac{d}{dt} \left\langle W, \Pi \right\rangle
= mg l \left\langle W, \boldsymbol{\Gamma} \times \boldsymbol{\chi}
\right\rangle
\]
where $ W (t) = \mathbf{R} (t) ^{-1} {\bf w} $; in other words, $W(t)$
is the body representation of a space fixed vector. This statement is
easily verified directly. Also, note that $ \left\langle W, \Pi
\right\rangle   =
\left\langle {\bf w}, \pi \right\rangle   $, with $\boldsymbol{\pi} =
\mathbf{R}(t)\boldsymbol{\Pi}$, so the Kelvin-Noether theorem may be
viewed as a statement about the rate of change of the momentum map of
the heavy top (that is, the spatial angular momentum) relative to the
{\it full group of rotations}, not just those about the vertical axis.

\section{General Lagrangian Reduction.}
The Lagrangian analogue of cotangent bundle reduction was developed
by Cendra, Ibort and Marsden [1987] and Marsden and Scheurle
[1993a,b]. (Routh, already around 1860 investigated in coordinates
what we would call today the abelian version.) These references
developed the coordinate version of the resulting reduced
Euler-Lagrange equations as well as the corresponding associated
reduction of variational principles. Another interesting paper in this
subject is that of Weinstein [1996] who used groupoids to show,
amongst other things, one can regard discrete and continuous Lagrangian
reduction from a single viewpoint. Discrete Lagrangian systems with
symmetry, following ideas of Veselov and Moser are of great interest in
numerical algorithms (see Wendlandt and Marsden [1997] and references
therein),

One starts with a $G$-invariant Lagrangian $L$ on $TQ$, which induces
a Lagrangian $l$ on the quotient space $(TQ)/G$, An important point,
but which is easy to see, is that, assuming the group actions are
free and proper, {\it this quotient space $(TQ)/G$ is intrinsically a
bundle over $T(Q/G)$ with a fiber modeled on the Lie algebra
$\mathfrak{g}$}. Denote, in a local trivialization, the variables in
the base, or {\it shape space} $Q/G$ by
$r ^\alpha , \dot{r} ^\alpha$  and the fiber variables by
$\xi ^\alpha$. In such a local trivialization, the reduced equations
are the {\bfi Hamel equations}:
\begin{eqnarray*}
\frac{d}{dt} \frac{ \partial l }{ \partial \dot{r} ^\alpha }
- \frac{ \partial l }{ \partial r ^\alpha } & = & 0 \\
\frac{d}{dt} \frac{ \partial l  }{ \partial \xi ^b } - \frac{
\partial l }{ \partial \xi ^a } C ^a _{ d b } \xi ^d & = & 0.
\end{eqnarray*}

However, for applications to stability problems as well as being
global, it is desirable to choose a connection $\mathcal{A}$, say
the mechanical connection, on the bundle $Q \rightarrow Q/G $. Then
the variables in the shape space and the vertical part $\Omega$ of a
velocity vector on $Q$ become globally and intrinsically defined.
In coordinates, the vertical part is given by
$\Omega^a=\xi^a+A^a_\alpha(r) \dot{r}^\alpha$
where the components of the connection
with respect to the local coordinates are denoted $A ^a_\alpha$. In
such variables,  the resulting {\bfi reduced Euler-Lagrange
equations} are
\begin{eqnarray*}
\frac{d}{dt} \frac{ \partial l }{ \partial \dot{r} ^\alpha }
- \frac{ \partial l }{ \partial r ^\alpha }
& = & \frac{ \partial l }{ \partial \Omega^a} ( - B ^a _{ \alpha
\beta } \dot{r} ^\beta + \mathcal{E} ^a _{ \alpha d } \Omega ^d ) \\
\frac{d}{dt} \frac{ \partial l  }{ \partial \Omega ^b } & = & \frac{ \partial l
}{ \partial \Omega ^a } (- \mathcal{E} ^a _{ \alpha \beta } \dot{r}
^\alpha + C ^a _{ d b } \Omega ^d )
\end{eqnarray*}
where $ B ^a _{ \alpha \beta } $ is the curvature of the connection
$\mathcal{A}^b _\alpha$, $ C^a _{bd}$ are the structure constants of the Lie
algebra $\mathfrak{g}$ and where $\mathcal{E} ^a_{\alpha d} =
C^a_{bd} \mathcal{A}^b _\alpha $. This second set of equations are
generalizations of the Euler-Poincar\'e equations. The two sets of
equations are called, respectively, the {\bfi horizontal\/} and {\bfi
vertical\/} equations.

This theory has had a large impact on, for example, the theory of
nonholonomic systems (see Bloch, Krishnaprasad, Marsden and
Murray [1996]).  It has also proven very useful in optimal control
problems. For example, it was used in Koon and Marsden [1997a] to
extend the falling cat theorem of Montgomery [1990] to the case of
nonholonomic systems.

The work of Cendra, Marsden and Ratiu [1997] develops the
geometry and variational structure of the reduced Euler-Lagrange
equations.  This work starts with the following view of the bundle
picture. As above, choose a connection
$\mathcal{A}$ on $ Q \rightarrow Q/G$ such as the mechanical connection and let
$\tilde{\mathfrak{g}}$ denote the vector bundle over shape space $Q / G$
which is the associated bundle to $\mathfrak{g}$ and the adjoint action of $G$
on $\mathfrak{g}$. There is a bundle isomorphism
\[ TQ/G \cong T(Q/G ) \oplus \tilde{ \mathfrak{g}}  \]
determined by the projection on the first factor and by the vertical projection
associated with the connection on the second. The sum is a Whitney sum (i.e.,
fiberwise a direct sum) of vector bundles over $Q/G$.

Using the geometry of this bundle $TQ/G = T(Q/G )\oplus
\tilde{\mathfrak{g}}$, one obtains a nice intrinsic interpretation of
the above reduced Euler-Lagrange equations in terms of covariant
derivatives of induced connections. One easily gets the dynamics of
particles in a Yang-Mills fields as well as many other interesting
examples (such as a rigid body with rotors, etc.) as special cases.
We will illustrate this with Wong's equations below. Preliminary
calculations show that in nonholonomic mechanics we will get a
beautiful geometric interpretation of the important {\it momentum
equation} of Bloch, Krishnaprasad, Marsden and Murray [1996].

\paragraph{The dual bundle picture.} The above construction
sheds light on the bundle picture for cotangent bundles
mentioned earlier. Taking the dual of the above isomorphism gives an
isomorphism
\[ T ^\ast Q/G \cong T ^\ast (Q/G ) \oplus \tilde{\mathfrak{g}} ^\ast . \]
Like any quotient Poisson manifold, the space $(T^\ast Q) / G $ has a
natural Poisson structure.  The description of this Poisson structure
viewed on the bundle $T ^\ast (Q/G ) \oplus
\tilde{\mathfrak{g}} ^\ast$  involves a synthesis of the canonical
bracket, the Lie-Poisson bracket, and curvature and is a very
interesting  way of encoding the original description of Montgomery,
Marsden and Ratiu [1984] of the Poisson structure on $(T ^\ast Q)/
G$. One gets from this picture {\it both} the usual Lie-Poisson
description of dynamics and the Hamiltonian description of
semidirect product theory.

\paragraph{Reduced variational principles.} The above picture also
leads naturally to a generalization of the reduced variational
principles described for the Euler-Poincar\'e equations above.
One does this by dividing the
variations in Hamilton's principle on $T^\ast Q$ into horizontal and
vertical parts, which then drop appropriately to the quotient space
in a manner similar to that in the Euler-Poincar\'e case.

We regard this reduction of variational principles as fundamental.
Another approach to reduction of Lagrangian systems is based on
reducing geometric objects such as almost tangent structures (see for
example, De Leon, Mello and Rodrigues [1992] and subsequent papers
of this group). The present framework seems to further
clarify this, and  identifies exactly which part of the reduced
space is a tangent bundle and which part is of Euler-Poincar\'e
type.

\paragraph{Stability under reduction.} One of the motivations is to
produce a category, a Lagrangian analogue of the symplectic and
Poisson category, that is stable under reduction. As the
Euler-Poincar\'e equations show, this category cannot be that of
tangent bundles or even locally tangent bundle-like objects.

However, the above ideas lead to an answer to this question in a
natural way. We enlarge the traditional category for
Lagrangian mechanics, namely tangent bundles, to bundles of the form
$$TQ\oplus U,$$ where $U$ is a vector bundle with a connection over
$Q$, each fiber of which carries a Lie algebra structure and where
the base $Q$ also carries a two form (a ``magnetic term''). This
structure is rich enough to include the reduced Euler-Lagrange
equations (by taking $ U = \tilde{\mathfrak{g}}$) and carry variational
methods.

In addition, {\it this category is stable under reduction}. If one
reduces this structure by the action of a group $K$, then $Q$ gets
replaced by $Q / K$, while the fiber expands by enlarging each Lie
algebra fiber by a group extension construction, with the
base two form playing the role of the cocycle. This
provides a Lagrangian analogue of the reduction by stages procedure
of Marsden, Misiolek, Perlmutter and Ratiu [1997].

\paragraph{Wong's Equations.}
As another illustration of the ideas of Lagrangian reduction, we
consider the important example of Wong's equations. See Montgomery
[1984] and the references therein for the original papers. This
example is important for its own sake, but also through the fact
that it enters into some fundamental optimal control problems, as
shown by Montgomery [1990] and Koon and Marsden [1997a].

We begin with a Riemannian manifold $Q$ with a free and proper
isometric action of a Lie group $G$ on $Q$. Let $ \mathcal{A} $
denote the mechanical connection; that is, it is the principal
connection whose horizontal space is the metric orthogonal to the
group orbits. The quotient space $ Q/G = X $ inherits a Riemannian
metric from that on  $Q$. Given a curve $ c (t) $ in $Q$, we shall
denote the corresponding curve in the base space $X$ by $r(t)$.

The optimal
control problem under consideration is as follows:
\vspace{0.2in}

\noindent {\bf Isoholonomic Problem (Falling Cat Problem).} Fixing
two points $ q _1 , \, q _2 \in Q $, among all curves
$q(t) \in Q, \, 0 \leq t \leq 1$ such that $q(0)=q_0, q(1)=q_1$ and
$\dot q(t)\in {\rm hor}_{q(t)}$ (horizontal with respect to the
mechanical connection $\mathcal{A}$),
find the curve or curves $q(t)$ such that the energy of the base
space curve, namely, \[ \frac{1}{2} \int_0^1 \|\dot {r} \| ^2 dt,\]
is minimized.

\begin{thm}[Montgomery (1990, 1991).]\label{optim2} If $q(t)$ is a
(regular) optimal trajectory for the isoholonomic problem, then
there exists a curve $\lambda (t) \in \mathfrak{g}^\ast$ such that
the reduced curve $ r(t) $ in $X = Q/G$ together with $ \lambda (t)
$ satisfies Wong's equations:
\begin{eqnarray*}
\dot p_{\alpha }& = & - \lambda _a\mathcal{B}  _{\alpha \beta}^a\dot
r^{\beta }-
\frac{1}{2}\frac
{\partial g^{\beta \gamma}}{\partial r^\alpha}p_{\beta
}p_{\gamma}\\
\dot \lambda _b& = & -\lambda _aC_{db}^a\mathcal{A}  _{\alpha}^d\dot
r^{\alpha}
\end{eqnarray*}
where $g_{\alpha \beta}$ is the local representation of the
metric on the base space $X$; that is
$$
 \frac{1}{2} \| \dot r \| ^2 = \frac{1}{2} g_{\alpha
\beta}\dot r^{\alpha}\dot r^{\beta},
$$
$g ^{\beta \gamma}$ is the inverse of the matrix $g_{\alpha\beta}$,
$p_{\alpha}$ is defined by
$$
p_{\alpha }= \frac{\partial l}{\partial \dot
r^{\alpha}}=g_{\alpha \beta}\dot r^\beta,
$$
and where we write the components of $ \mathcal{A}$ as
$ \mathcal{A}  ^b _\alpha $ and similarly for its curvature $
\mathcal{B} $.
\end{thm}

\noindent{\bf Proof.\,} By general principles in the calculus of
variations, given an optimal solution $ q (t) $, there is a Lagrange
multiplier
$\lambda (t)$ such that the new action function defined on the space
of curves with fixed endpoints by
\[
\mathfrak{S} [q ({} \cdot{} )] =  \int _0 ^1 \left[  \frac{1}{2} \|
\dot r(t)  \| ^2 + \left\langle \lambda (t) , \mathcal{A} \dot{q}(t)
\right\rangle \right] dt
\]
has a critical point at this curve. Using the integrand as a
Lagrangian, identifying $ \Omega = \mathcal{A} \dot{q}$ and applying
the reduced Euler-Lagrange equations to the reduced Lagrangian
\[ l ( r, \dot{r} , \Omega ) = \frac{1}{2} \| \dot{r} \| ^2 +
\left\langle \lambda , \Omega \right\rangle  \]
then
gives Wong's equations by the following simple calculations:
\[
\frac{\partial l}{\partial \dot r^{\alpha}}=
g_{\alpha \beta}\dot r^\beta;\quad
\frac{\partial l}{\partial r^{\alpha}}=
\frac {1}{2}{\frac {\partial g^{\beta \gamma}}{\partial r
^{\alpha}}}\dot r^{\beta}\dot r^{\gamma};\quad
\frac {\partial l}{\partial \Omega ^a}= \lambda _a.
\]
The constraints are $ \Omega = 0 $ and so the reduced Euler-
Lagrange equations become
\begin{eqnarray*}
\frac{d}{dt}\frac{\partial l}{\partial \dot r^{\alpha}}-
\frac{\partial l}{\partial r^{\alpha}}& = & -\lambda _a(\mathcal{B}
_{\alpha
\beta }^a\dot r^\beta)\\
\frac{d}{dt}\lambda _b& = & -\lambda _a({\cal E} _{\alpha b}^a\dot
r^\alpha)=-\lambda _aC_{db}^a\mathcal{A}  _{\alpha}^d\dot r^{\alpha}.
\end{eqnarray*}
But
\begin{eqnarray*}
\frac {d}{dt}\frac{\partial l}{\partial \dot r^{\alpha}}-
\frac{\partial l}{\partial r^{\alpha}}&=&\dot p_{\alpha}-
\frac {1}{2}{\frac {\partial g_{\beta \gamma}}{\partial r
^{\alpha}}}\dot r^{\beta}\dot r^{\gamma}\\
&=&\dot p_{\alpha}+\frac{1}{2}{\frac {\partial g^{\kappa
\sigma}}{\partial r ^{\alpha}}}g_{\kappa \beta}g_{\sigma
\gamma}\dot r^{\beta}\dot r^{\gamma}\\
&=& \dot p_{\alpha}+\frac{1}{2}{\frac {\partial g^{\beta
\gamma}}{\partial r ^{\alpha}}}p_{\beta}p_{\gamma},\\
\end{eqnarray*}
\noindent
and so we have the desired equations. \quad $\blacksquare$
\bigskip

Let us now give another version of this procedure in abstract
notation and, at the same time, show how it fits in with the bundle
view of Lagrangian reduction. Again, let $\pi :Q \rightarrow X$ be a
principal bundle with structure group $G$, a Lie group acting on the
left, let $\mathcal{A} $ be a principal connection on $Q$ and let
$\mathcal{B}$ be the curvature of $\mathcal{A} $. Suppose that $g$
is a given Riemannian metric on $X$ and let $\nabla$  be the
corresponding Levi-Civita connection.  Assume, for simplicity, that
$G$ is a compact group with a bi-invariant Riemannian metric $K$.

Define the Lagrangian $L : TQ \rightarrow {\mathbb R}$ by
\[
L(q, \dot{q}) = \frac{1}{2}K\left(A(q, \dot{q}), A(q, \dot{q})
\right) + \frac{1}{2}g\left(\pi (q)\right)
\left(T\pi (q, \dot{q}), T\pi
(q,
\dot{q})\right).
\]
An element of $\tilde{\mathfrak{g}}$ has the form
$\bar{v} = [q, v]_G$ where $q \in Q$ and $v \in \mathfrak{g}$ and
where $[q, v]_G$ denotes the equivalence class of the pair $(q,v) $
with respect to the group action by $G$.  Since
$K$ is bi-invariant, its restriction to
$\mathfrak{g}$ is ${\rm Ad}$-invariant, and therefore we can define
$K\left([q, v]_G, [q, v]_G\right) =
K(v, v)$.

The reduced bundle is $$T(Q/G)\oplus \tilde{\mathfrak{g}} \equiv
TX\oplus \tilde{\mathfrak{g}}$$ and a typical element of it is
denoted
$(x, \dot{x}, \bar{v})$. The reduced Lagrangian is given by
\[
l (x, \dot{x}, \bar{v}) =
\frac{1}{2}K(v, v)
+ \frac{1}{2}g(x)(x, \dot{x})\,.
\]
Now we will write the vertical and
horizontal reduced Euler-Lagrange equations.  An arbitrary variation
$\delta \bar{v}$ in the direction
of the fiber of $\tilde{\mathfrak{g}}$ is of type
$\delta \bar{v} = [q, \delta v]_G$, where $\delta v \in
\mathfrak{g}$ is arbitrary. We have
\[
\frac{\partial l}{\partial \bar{v}}(x, \dot{x}, \bar{v})
\delta \bar{v} = K(v, \delta v)
\]
and
\[
\left({\rm ad}^{\ast}_{\bar{v}}\frac{\partial l}{\partial \bar{v}}
(x, \dot{x}, \bar{v})\right)\delta \bar{v} = K\left(v, [v, \delta
v]\right)\,.
\]
Since $K$ is bi-invariant, we have
$K({\rm ad}_w u, v) +  K(u, {\rm ad}_w v) = 0,$
 for all $u$, $v$, and $w$ in $\mathfrak{g}$.
Therefore,
$K\left(v, [v, \delta v]\right) = -K\left([v, v], \delta v\right) =
0$.
Thus the vertical reduced Euler-Lagrange equation is one of Wong's
equations
\[
\frac{d}{dt}K(v, {} \cdot {}) = 0.
\]
This actually agrees with the previous form of Wong's equations
since $\lambda$ represents the value of the {\it body} momentum
while the form $K(v, {} \cdot {})$ represents the {\it spatial}
momentum, which is conserved.  After some  straightforward
calculations,  the horizontal reduced Euler-Lagrange equation
becomes  in this case
\[
-\left(\nabla _{\dot{\gamma}}\dot{\gamma}\right)^{\flat} =
K\left(v, \tilde{\mathcal{B}}(x)(\dot{x}, {} \cdot {})\right)\,,
\]
which is the other of Wong's equations.
Here $ \tilde{ \mathcal{B} } $ denotes the curvature thought of as a
Lie algebra valued two form on the base.

\section{The Euler-Poincar\'e Equations via Lagrangian Reduction}

Now we are ready to explain how the Euler-Poincar\'e equations can
be derived using the techniques of Lagrangian reduction. This should be
plausible from what we have already said simply because the
coordinate expression for the reduced Euler-Lagrange equations include
a generalized form of Euler-Poincar\'e equations. But we need to make
this connection precise.

\paragraph{The set up.} We consider a slight generalization of the
Euler-Poincar\'e equations given earlier. This time we consider a
group $G$ acting on a vector space $V$ and hence on its dual
$V^\ast$. We also consider a configuration space $Q$, but in this
section, $ G $ will act trivially on $Q$.  Consider a Lagrangian
\[
L : T(G\times Q) \times V^\ast \rightarrow  {\mathbb R}
\]
where $G$ is a group, $Q$ is a manifold and $V^\ast $ is the dual of the
vector space $V$. The value of $L$ at the point
$(g, q, \dot{g}, \dot{q}, a_0) \in T(G\times Q) \times V^\ast$
will
be denoted
$L(g, q, \dot{g}, \dot{q}, a_0)$,
as usual, and, as in the discussion of the Euler-Poincar\'e
equations earlier, we think of $a_0$ as being a parameter that
remains  fixed along the evolution of the system.

By construction, the action of $G$ on $V$ and $V^\ast$ satisfies the
property
$$\left\langle ga, gv \right\rangle = \left\langle  a, v
\right\rangle$$  for all
$a \in V^\ast$, all $v \in V$ and all $g \in G$.
(We will often write $\left\langle  a, v \right\rangle  =
\left\langle  v, a \right\rangle  $).

Assume that $L$ has the
following  invariance property:
\[
L(g^\prime g, q, g^\prime \dot{g}, \dot{q}, g^\prime a_0) = L(g, q, \dot{g},
\dot{q}, a_0)
\]
for all
$a_0 \in V^\ast$, all $q \in Q$ and all $g^\prime, g \in G$.
Let
\[
L(e, q, \xi, \dot{q}, a) = l (\xi, q, \dot{q}, a),
\]
for all $\xi \in \mathfrak{g}$,  all $q \in Q$ and all
$a \in V^\ast$. Then the invariance property implies,
for all $g \in G$,  all $q \in Q$ and all
$a \in V^\ast$,
\[
L(g, q, \dot{g}, \dot{q}, a_0) = l (\xi, q, \dot{q}, a)
\]
where $\xi = g^{-1}\dot{g}$ and $a = g^{-1}a_0$.

A small generalization of the argument given earlier for the
Euler-Poincar\'e equations proves the next result.

\begin{thm} The following
conditions are equivalent:
\begin{itemize}
\item[(i)]
The curve $\left(g(t), q(t), a_0\right)$ is a critical point of the action
\[
\int _{t_0}^{t_1}L(g, q, \dot{g}, \dot{q}, a_0) dt
\]
with restrictions on variations given by $\delta g(t_i) = 0$ for
$i = 0, 1$, $\delta q(t_i) = 0$ for
$i = 0, 1$, and $\delta a_0 = 0$.
\item[(ii)]
The curve $\left(\xi (t),  q(t), a(t)\right)$,
where $a(t) = g^{-1}(t)a_0$ for all $t$
and  $\xi (t) = g^{-1}(t)\dot{g}$,
is a  critical point of the action
\[
\int _{t_0}^{t_1}l (\xi, q, \dot{q}, a) dt
\]
with restrictions on variations given by
\[
\delta \xi = \dot{\eta} + [\xi, \eta]
\]
where $\eta$ is any curve on
$\mathfrak{g}$ such that $\eta (t_i) = 0$ for
$i = 0, 1$,
\[
\delta q(t_i) = 0
\]
for
$i = 0, 1$ and
\[
\delta a = - \eta a
\]
and, besides, the curve $a(t)$ must
satisfy
\[
\dot{a} + \xi a = 0
\]
for all $t$.
(As earlier, this last condition comes from the condition
$\dot{a}_0 = 0$ together with $a_0 = ga$.)
\end{itemize}
\end{thm}

A direct application of $(ii)$ leads to the
reduced equations
\[
\frac{d}{dt}\frac{\delta   l}{\delta  \xi}
= {\rm ad}_{\xi}^{\ast}
\frac{\delta  l}{\delta \xi} +
\frac{\delta l}{\delta a} \diamond  a
\]
together with the standard Euler-Lagrange equations for $q$:
\[
\frac{\partial l}{\partial q} -
\frac{d}{dt}\frac{\partial l}{\partial \dot{q}} = 0.
\]
Recall that, by definition,
\[
\left\langle v \diamond a, \zeta\right\rangle  = - \left\langle \zeta  a, v
\right\rangle
\]
for all $\zeta  \in \mathfrak{g}$, all $a \in V^\ast$ and all
$v
\in V$.
The first of these equations is of course the {\bfi Euler-Poincar\'e}
equation. These and the Euler-Lagrange equations for $q$, together
with the equation $\dot{a} + \xi a = 0$  form a complete set of
equations of the system in terms of the variables
$(\xi, q, a)$. This framework is applied to Vlasov plasmas in Cendra, Holm,
Hoyle
and Marsden [1997].

Now we shall recast conditions $(i)$ and $(ii)$ into an equivalent
form.  The idea is to introduce the condition that $a_0$ is conserved by
making it the momentum conjugate to a cyclic variable, just as one does for
the charge in Kaluza-Klein theory.  Thus, let us define the Lagrangian
\[
\bar{L} : T(G\times Q \times V^{\ast} \times V) \rightarrow  {\mathbb R}
\]
by
\[
\bar{L} (g, q, a_0, v_0 , \dot{g}, \dot{q}, \dot{a}_0, \dot{v}_0 ) =
L(g, q, \dot{g}, \dot{q}, a_0 ) + \left\langle
a_0, \dot{v}_0 \right\rangle .
\]
Notice that for the Lagrangian $ \bar{L} $, $ a _0 $ is the momentum
conjugate to the cyclic variable $v_0$, and hence is a constant. The
variable $v_0$ is not constant, but its evolution is the first order
equation $\dot{v}_0  + \partial L /\partial a _0  = 0 $.
Thus, the Euler-Lagrange equations for
$\bar{L}$ are equivalent to the Euler-Lagrange equations for $L$
with the parameter $ a _0 $ fixed (together with the equation for
$v_0$).

Thus, the strategy is to perform reduction of $L$ by the equivalent process
of performing standard tangent bundle La\-gran\-gian reduction of $\bar{
L }$.

Now we observe that
\[ G\times Q \times V^{\ast} \times V \rightarrow (G\times Q \times V^{\ast}
\times V)/G
\] is  a principal
bundle with structure group $G$ acting as before, that is,
$g^\prime (g, q, a_0, v_0) = (g^\prime g, q, g^\prime a_0, g^\prime v_0)$.
Moreover,
this bundle is
isomorphic, as a principal bundle, to  the trivial bundle
\[
G\times Q \times V^{\ast} \times V \rightarrow Q \times V^{\ast} \times V
\] where the action of $G$ is given by
$g^\prime\cdot(g, q, a, v) = (g^\prime g, q, a, v)$, for all $g^\prime, g
\in G$, all
$a \in V^\ast$ and all $v \in V$. This assertion follows using
the isomorphism
$\psi : G\times Q \times V^{\ast} \times V \rightarrow  G\times Q \times
V^{\ast} \times V$
given by $\psi (g, q, a_0, v_0) = (g, q, g^{-1}a_0, g^{-1}v_0)
\equiv (g, q, a, v)$. It is easy to check that
$\psi \left(g^\prime(g, q, a_0, v_0)\right) = g^\prime\cdot \psi (g, q, a_0,
v_0)
\equiv g^\prime \cdot (g, q, a, v)$,  for all $g^\prime , g \in G$, all
$a_0 \in V^\ast$ and all $v_0 \in V$.
One also checks that the composition
$\bar{L}\circ T\psi ^{-1} = : L^V$ is given by
\[
L^V (g, q, a, v, \dot{g}, \dot{q}, \dot{a}, \dot{v}) = L(g, q,\dot{g},
\dot{q}, a)  + \left\langle  a, \dot{v} + g^{-1}\dot{g}v \right\rangle .
\]
>From now on we will use the trivial bundle described above and
the Lagrangian  $L^V$.

\noindent {\bf Remark.} Using techniques like those in Cendra and Marsden
[1987], which give a version  of the Lagrange multiplier theorem, we can
show that conditions $(i)$ and $(ii)$ are equivalent to either of the
following conditions,
{\it
\begin{itemize}
\item[(iii)]
The curve $(g(t), q(t), a_0, v_0)$ is a critical point of the action
\[
\int _{t_0}^{t_1} \bar{L} (g, q, a_0, v_0, \dot{g}, \dot{q}, \dot{a}_0,
\dot{v}_0) dt
\]
for variations satisfying the endpoint conditions
$\delta g(t_i) = 0$ ,
$\delta q(t_i) = 0$ ,
$\delta a_0(t_i) = 0$ and
$\delta v_0(t_i) = 0$ for $i = 0, 1$.
\item[(iv)]
The curve $\left(g(t), q(t), a(t), v(t)\right)$ is a critical point
of the action
\[
\int _{t_0}^{t_1} L^V (g, q, a, v, \dot{g}, \dot{q}, \dot{a}, \dot{v})
dt
\]
for variations satisfying the endpoint conditions
$\delta g(t_i) = 0$, $\delta q(t_i) = 0$,
$\delta a(t_i) = 0$ and $\delta v(t_i) = 0$ for $i = 0, 1$. \quad
$\blacklozenge$
\end{itemize}
}
We shall show that the reduced Euler-Lagrange equations for the
Lagrangian $ L ^V $ are the same as the reduced system for $l$.
To do this, we first calculate the reduced bundle $T(Q\times V^\ast \times
V)\oplus
\tilde{\mathfrak{g}}$.
Since the bundle $ G \times Q \times V^\ast\times V $ is trivial, the bundle
$\tilde{\mathfrak{g}}$
is
$\tilde{\mathfrak{g}} \equiv Q\times V^\ast\times V \times \mathfrak{g}$.
The Lie algebra structure on
$\tilde{\mathfrak{g}}$ is given by
$\left[(q, a, v, \xi _1), (q, a, v, \xi _2)\right] =
(q, a, v, [\xi _1, \xi _2])$.

Let us choose the trivial principal connection
$\mathcal{A}$ on
$G\times Q \times V^{\ast} \times V $, that is, the connection given by
$\mathcal{A} (g, q, a, v, \dot{g}, \dot{q}, \dot{a}, \dot{v}) =
g^{-1}\dot{g}$.  Since the connection is trivial, it has zero curvature.
The reduced Lagrangian is clearly given by
\[
l^V (\xi , q, a, v,  \dot{q}, \dot{a}, \dot{v}) = l(\xi , q,
\dot{q}, a)  + \left\langle  a, \dot{v} + \xi v \right\rangle .
\]

Now one is ready to calculate the reduced Euler-Lagrange equations for $ l
^V $ using this set up. This can be done in coordinates using the
coordinate version of the reduced Euler-Lagrange equations we gave
earlier, or in their intrinsic formulation.

Since the connection is trivial and its curvature is zero, the horizontal
equations for $l ^V$  are simply the usual Euler-Lagrange equations for
these variables; these give the equations for $q$ as well as the equations
for $a$ and $v$. The Euler-Lagrange equation for $q$ with respect to $ l^V
$ is clearly the same as the Euler-Lagrange equation for $q$ with respect
to $l$. The horizontal equation for $a$ is the equation
\[ \frac{d}{dt} \frac{ \partial l^V}{ \partial \dot{a} } - \frac{ \partial
l^V }{  \partial a } = 0 \]
which, since $ l^V $ does not depend on $\dot{a}$, is simply
\[ \dot{v} + \xi v + \frac{ \partial l }{ \partial a } = 0, \]
which  is equivalent to the
equation $\dot{v}_0  + \partial L / \partial a _0 = 0 $ that we had before.

The Euler-Lagrange equation for $v$ is
\[ \frac{d}{dt} \frac{ \partial l^V}{ \partial \dot{v} } - \frac{ \partial
l^V }{  \partial v } = 0. \]
Since $ \partial l^V / \partial \dot{v} = a $ and $ \partial l^V / \partial
v = - \xi a $, this gives the correct equation $ \dot{a} + \xi a = 0 $ for
$a$.

The vertical equation becomes
\[
\frac{d}{dt} \frac{  \partial l^V }{ \partial \xi }
= {\rm ad} ^\ast _{ \xi }
\left( \frac{ \partial l^V }{ \partial \xi } \right).
\]
Clearly
\[ \frac{  \partial l^V }{ \partial \xi } = \frac{\partial l}{\partial
\xi } + v \diamond a , \]
and so the vertical equation becomes
\[
\frac{d}{dt}  \frac{  \partial l }{ \partial
\xi } + \dot{v}  \diamond a  + v \diamond \dot{a}  = {\rm ad} ^\ast _{ \xi }
 \frac{  \partial l }{ \partial
\xi } +  {\rm ad} ^\ast _{ \xi }(v \diamond a)  .
\]
Using the equations for $ \dot{a} $ and $ \dot{v} $, this becomes
\[
\frac{d}{dt}  \frac{  \partial l }{ \partial
\xi } -  \left( \xi v + \frac{ \partial l }{ \partial a } \right)  \diamond
a  -  v
\diamond (\xi a)  = {\rm ad} ^\ast _{
\xi }
 \frac{  \partial l }{ \partial
\xi } +  {\rm ad} ^\ast _{ \xi }(v \diamond a)  .
\]
i.e.,
\[
\frac{d}{dt}  \frac{  \partial l }{ \partial
\xi }  - {\rm ad} ^\ast _{\xi } \frac{  \partial l }{ \partial
\xi } - \frac{ \partial l }{ \partial a } \diamond a =  (\xi v )
\diamond a  +  v
\diamond (\xi a)  +  {\rm ad} ^\ast _{ \xi }(v \diamond a)  .
\]
But the right hand side is identically zero, as is seen from the
definitions. Thus, we have achieved our goal of recovering the
Euler-Poincar\'e equations with an advected parameter from Lagrangian
reduction.

%refmark

\section*{References}
\begin{description}

\item Abarbanel, H.D.I., D.D. Holm, J.E. Marsden, and T.S.
Ratiu [1986]  Nonlinear stability analysis of stratified
fluid equilibria, {\it Phil. Trans. Roy. Soc. London A\/}
{\bf 318}, 349--409; also Richardson number criterion for
the nonlinear stability of three-dimensional stratified
flow, {\it Phys. Rev. Lett.\/} {\bf 52} [1984], 2552--2555.

\item Abraham, R. and J.E. Marsden [1978]
{\it Foundations of Mechanics\/}, Second Edition,
Addison-Wesley.

\item Arnold, V.I. [1966a]
Sur la g\'{e}ometrie differentielle
des groupes de Lie de dimenson
infinie et ses applications \`{a}
l'hydrodynamique des fluids parfaits,
{\it Ann. Inst. Fourier, Grenoble\/} {\bf 16}, 319--361.

\item Arnold, V.I. [1966b]
On an a priori estimate in the theory of
hydrodynamical stability,
{\it Izv. Vyssh. Uchebn. Zaved. Mat. Nauk\/}
{\bf 54}, 3--5; English
Translation: {\it Amer. Math. Soc. Transl.\/}
{\bf 79} [1969], 267--269.

\item Arnold, V.I. [1966c]
Sur un principe variationnel pour les d\'ecoulements
stationaires des liquides parfaits et ses
applications aux problemes de stabilit\'e non lin\'eaires,
{\it  J. M\'ecanique\/} {\bf 5}, 29--43.

\item Arnold, V.I. (ed.) [1988]
{\it Dynamical Systems III\/},
Encyclopedia of Mathematics {\bf 3}, Springer-Verlag.

\item Arnold, V.I. and B. Khesin [1992]
Topological methods in hydrodynamics,
{\it Ann. Rev. Fluid Mech.\/} {\bf 24}, 145--166.

\item Arnold, V.I. and B. Khesin [1997]
{\it Topological methods in Fluid Dynamics},
Appl. Math. Sciences, Springer-Verlag.

\item Bloch, A.M., P.S. Krishnaprasad, J.E. Marsden, and R. Murray
[1996] Nonholonomic mechanical systems with symmetry,
{\it Arch. Rat. Mech. An.}, {\bf 136}, 21--99.

\item Bloch, A.M., P.S. Krishnaprasad,
J.E. Marsden, and  T.S. Ratiu [1994]
Dissipation induced instabilities,
{\it Ann. Inst. H. Poincar\'{e}, Analyse Nonlineare\/}
{\bf 11}, 37--90.

\item Bloch, A.M., P.S. Krishnaprasad,
J.E. Marsden, and T.S. Ratiu [1996]
The Euler-Poincar\'{e} equations and
double bracket dissipation,
{\it Comm. Math. Phys.}  {\bf 175}, 1--42.

\item Cendra, H. and J.E. Marsden [1987]
Lin constraints, Clebsch potentials and variational principles,
{\it Physica D\/} {\bf 27}, 63--89.

\item Cendra, H., A. Ibort, and J.E. Marsden [1987]
Variational principal fiber bundles: a geometric
theory of Clebsch potentials and Lin constraints,
{\it J.  Geom.  Phys.\/} {\bf 4}, 183--206.

\item  Cendra, H., D.D. Holm,
M.J.W. Hoyle and J. E. Marsden [1997]
The Maxwell-Vlasov equations in Euler-Poincar\'{e} form,
{\it preprint.}

\item  Cendra, H.,  J. E. Marsden and T.S. Ratiu [1997]
Lagrangian reduction by stages, {\it preprint.}

\item Chetayev, N.G. [1941]
On the equations of Poincar\'{e},
{\it J. Appl. Math. Mech.\/} {\bf 5}, 253--262

\item De Leon, M. M. H. Mello, and P.R. Rodrigues [1992]
Reduction of nondegenerate nonautonomous Lagrangians.
{\it Cont. Math. AMS\/} {\bf 132}, 275-306.

\item Ebin, D.G. and J.E. Marsden [1970]
Groups of diffeomorphisms and the motion of an incompressible
fluid, {\it Ann. Math.\/} {\bf 92}, 102--163.

\item Guichardet, A. [1984]
On rotation and vibration motions of molecules,
{\it Ann. Inst. H. Poincar\'{e} \/} {\bf 40}, 329--342.

\item Guillemin, V. and S. Sternberg [1980]
The moment map and collective motion,
{\it Ann. of Phys.\/} {\bf 1278}, 220--253.

\item Guillemin, V. and S. Sternberg [1984]
{\it Symplectic Techniques in Physics\/},
Cambridge University Press.

\item Hamel, G [1904]
Die Lagrange-Eulerschen Gleichungen der Mechanik,
{\it Z. f\"{u}r Mathematik u. Physik\/} {\bf 50}, 1--57.

\item Hamel, G [1949]
{\it Theoretische Mechanik\/}, Springer-Verlag.

\item Holm, D.D. [1996]
Hamiltonian balance equations,
{\it Physica D}, {\bf 98} (1996) 379-414.

\item Holm, D.D. and B.A. Kupershmidt [1983]
Poisson brackets and Clebsch representations for
magnetohydrodynamics, multifluid plasmas, and elasticity,
{\it Physica D\/} {\bf 6}, 347--363.

\item Holm, D.D., J.E. Marsden, and T.S. Ratiu [1986]
The Hamiltonian structure of continuum mechanics in
material, spatial and convective representations,
{\it S\'{e}minaire de Math\'{e}matiques sup\'{e}rie,
Les Presses de L'Univ. de Montr\'{e}al\/} {\bf 100}, 11--122.

\item Holm, D. D., Marsden, J. E. and Ratiu, T. [1997]
The Euler-Poincar\'{e} equations and semidirect products with
applications to continuum theories, {\it preprint.}

\item Holm, D.D., J.E. Marsden, T.S. Ratiu, and A. Weinstein [1985]
Nonlinear stability of fluid and plasma equilibria,
{\it Phys. Rep.\/} {\bf 123}, 1--116.

\item Holmes, P.J. and J.E. Marsden [1983]
Horseshoes and Arnold diffusion for Hamiltonian
systems on Lie groups,
{\it Indiana Univ. Math. J.\/} {\bf 32}, 273--310.

\item Iwai, T. [1987]
A geometric setting for classical molecular dynamics,
{\it Ann. Inst. Henri Poincar\'{e}, Phys. Th.\/}
{\bf 47}, 199--219.

\item Koon, W.S. and J.E. Marsden [1997a] Optimal control for
holonomic and nonholonomic mechanical systems with symmetry and
Lagrangian reduction,
{\it SIAM J. Control and Optim.\/} {\bf 35}, 901--929.

\item Kummer, M. [1981]
On the construction of the reduced phase space of a
Hamiltonian system with symmetry,
{\it Indiana Univ. Math. J.\/} {\bf 30}, 281--291.

\item Kupershmidt, B.A. and T. Ratiu [1983] Canonical maps
between semidirect products with applications to elasticity and
superfluids,
{\it Comm. Math. Phys.\/} {\bf 90}, 235--250.

\item Lagrange, J.L. [1788]
{\it M\"{e}canique Analitique\/},
Chez la Veuve Desaint

\item Leonard, N.E. and J.E. Marsden [1997]
Stability and Drift of Underwater Vehicle Dynamics:
Mechanical Systems with Rigid Motion Symmetry,
{\it Physica D\/} {\bf 105}, 130--162.

\item Lie, S. [1890]
{\it Theorie der Transformationsgruppen, Zweiter Abschnitt\/},
Teubner, Leipzig.

\item Marsden, J.E. [1982]
A group theoretic approach to the equations of plasma physics,
{\it Can. Math. Bull.\/} {\bf 25}, 129--142.

\item Marsden, J.E., G. Misiolek, M. Perlmutter and T.S. Ratiu [1997]
Reduction by stages and group extensions, {\it in preparation\/}.

\item Marsden, J.E., G.W. Patrick and S. Shkoller [1997]
Variational methods in continuous and discrete
mechanics and field theory, {\it preprint}.

\item Marsden, J.E. and T.S. Ratiu [1994]  {\it Introduction to
Mechanics and Symmetry\/}, Texts in Applied Mathematics, {\bf  17},
Springer-Verlag.

\item Marsden, J.E., T.S. Ratiu, and A. Weinstein [1984a]
Semi-direct products and reduction in mechanics,
{\it Trans. Am. Math. Soc.\/} {\bf 281}, 147--177.

\item Marsden, J.E., T.S. Ratiu, and A. Weinstein [1984b]
Reduction and Hamiltonian structures on
duals of semidirect product Lie Algebras,
{\it Cont. Math. AMS\/} {\bf 28}, 55--100.

\item Marsden, J.E. and J. Scheurle [1993a]
Lagrangian reduction and the double spherical pendulum,
{\it ZAMP\/} {\bf 44}, 17--43.

\item Marsden, J.E. and J. Scheurle [1993b]
The reduced Euler-Lagrange equations,
{\it Fields Institute Comm.\/} {\bf 1}, 139--164.

\item Marsden, J.E. and A. Weinstein [1974]  Reduction of
symplectic manifolds with symmetry,
{\it Rep. Math. Phys.\/} {\bf 5}, 121--130.

\item Marsden, J.E., A. Weinstein, T.S. Ratiu,
R. Schmid, and R.G. Spencer [1983]
Hamiltonian systems with symmetry, coadjoint
orbits and plasma physics, in
Proc. IUTAM-IS1MM Symposium on
{\it Modern Developments in Analytical Mechanics\/},
Torino 1982, {\it Atti della Acad. della Sc. di Torino\/}
{\bf 117}, 289--340.

\item Montgomery, R. [1984]
Canonical formulations of a particle in a Yang-Mills field,
{\it Lett. Math. Phys.\/} {\bf 8}, 59--67.

\item Montgomery, R. [1986] The bundle picture in mechanics , Ph.D.
Thesis, Berkeley.

\item Montgomery, R. [1990]
Isoholonomic problems and some applications,
{\it Comm. Math Phys.\/} {\bf 128}, 565--592.

\item Montgomery, R. [1991]
Optimal control of deformable bodies and its
relation to gauge theory, in
{\it The Geometry of Hamiltonian Systems}, T. Ratiu ed.,
Springer-Verlag.

\item Montgomery, R., J.E. Marsden, and T.S. Ratiu [1984]
Gauged Lie-Poisson structures,
{\it Cont. Math. AMS\/} {\bf 28}, 101--114.

\item Ovsienko, V.Y. and B.A. Khesin [1987]
Korteweg-de Vries superequations as an Euler equation,
{\it Funct. Anal. and Appl.\/} {\bf 21}, 329--331.

\item Poincar\'{e}, H. [1901] Sur la stabilit\'{e} de
l'\'{e}quilibre  des figures piriformes affect\'{e}es par une
masse fluide en rotation,
{\it Philosophical Transactions A\/} {\bf 198}, 333--373.

\item Poincar\'{e}, H. [1910] Sur la precession des corps
deformables,
{\it Bull Astron\/} {\bf 27}, 321--356.

\item Ratiu. T.S.  [1980a] The Euler-Poissot equations and
integrability {\it Thesis\/}, University of California at Berkeley.

\item Ratiu. T.S. [1981]  Euler-Poisson equations on Lie
algebras and the $N$-dimensional heavy rigid body,
{\it Proc. Natl. Acad. Sci. USA \/} {\bf 78}, 1327--1328.

\item Ratiu, T.S. [1982]  Euler-Poisson equations on Lie
algebras and the $N$-dimensional heavy rigid body,
{\it Am. J. Math.\/} {\bf 104}, 409--448, 1337.

\item Satzer, W.J. [1977]
Canonical reduction of mechanical systems
invariant under abelian group actions
with an application to celestial mechanics,
{\it Ind. Univ. Math. J.\/} {\bf 26},
951--976.

\item Smale, S. [1970]
Topology and mechanics,
{\it Inv. Math.\/} {\bf 10}, 305--331, {\bf 11}, 45--64.

\item Sudarshan, E.C.G. and N. Mukunda [1974]
{\it Classical Mechanics: A Modern Perspective\/},
Wiley, New York, 1974; Second Edition, Krieber,
Melbourne--Florida, 1983.

\item Vinogradov, A.M. and B.A. Kupershmidt [1977]
The structures of Hamiltonian mechanics,
{\it Russ. Math. Surv.\/} {\bf 32}, 177--243.

\item Weinstein, A. [1996]
Lagrangian Mechanics and Groupoids
{\it Fields Inst. Comm.\/}, {\bf 7}, 207--231.

\item Wendlandt, J.M. and J.E. Marsden [1997]  Mechanical
integrators derived from a discrete variational principle,  {\it
Physica D\/} {\bf  106}, 223--246.

\end{description}

\end{document}